\documentclass[fleqn,usenatbib]{mnras}

\usepackage{newtxtext,newtxmath}

\usepackage[T1]{fontenc}

\DeclareRobustCommand{\VAN}[3]{#2}
\let\VANthebibliography\thebibliography
\def\thebibliography{\DeclareRobustCommand{\VAN}[3]{##3}\VANthebibliography}

\usepackage{graphicx}	
\usepackage{amsmath}	

\outer\def\gtae {$\buildrel {\lower3pt\hbox{$>$}} \over 
{\lower2pt\hbox{$\sim$}} $}
\outer\def\ltae {$\buildrel {\lower3pt\hbox{$<$}} \over 
{\lower2pt\hbox{$\sim$}} $}
\newcommand{\Msun}{$M_{\odot}$}

\newcommand{\Rsun}{$R_{\odot}$}

\newcommand{\tess}{\it TESS}

\newcommand{\rchi}{$\chi^{2}_{\nu}$}
\newcommand{\chisq}{$\chi^{2}$}

\newcommand{\locccf}{CCF$_{\rm{loc}}$ }
\newcommand{\DIccf}{CCF$_{\rm{DI}}$ }
\newcommand{\locccfs}{CCFs$_{\rm{loc}}$ }
\newcommand{\DIccfs}{CCFs$_{\rm{DI}}$ }

\DeclareUnicodeCharacter{2212}{-}

\newcommand{\Nperiod}{\mbox{$5.3220125 \pm 0.0000010$}}
\newcommand{\Ntc}{\mbox{$2458282.25880 \pm 0.00019$}}  

\newcommand{\Nars}{\mbox{$8.37 \pm 0.15$}}
\newcommand{\Ninc}{\mbox{$84.96 \pm 0.17$}}
\newcommand{\Nuonets}{\mbox{$0.449\substack{+0.009 \\ -0.009}$}}
\newcommand{\Nutwots}{\mbox{$0.230\substack{+0.013 \\ -0.013}$}}
\newcommand{\Naau}{\mbox{$0.0608 \pm 0.0013$}}
\newcommand{\Nduration}{\mbox{$3.845 \pm 0.020$}}  
\newcommand{\Nimpact}{\mbox{$0.735 \pm 0.011$}}
\newcommand{\Nmstar}{\mbox{$1.060 \pm 0.068$}}
\newcommand{\Nrstar}{\mbox{$1.561 \pm 0.043$}}
\newcommand{\Nloggstar}{\mbox{$4.075 \pm 0.018$}}
\newcommand{\Nrhostar}{\mbox{$0.278 \pm 0.016$}}
\newcommand{\Nmplanet}{\mbox{$0.273 \pm 0.019$}}
\newcommand{\Nrplanet}{\mbox{$1.226 \pm 0.037$}}  
\newcommand{\Nloggplanet}{\mbox{$2.619 \pm 0.031$}}
\newcommand{\Nrhoplanet}{\mbox{$0.148 \pm 0.013$}}

\newcommand{\Nrprssq}{\mbox{$0.006517 \pm 0.000061$}}
\newcommand{\Nreflex}{\mbox{$30.5 \pm 1.7$}}
\newcommand{\Ngamma}{\mbox{$-19\,661.5 \pm 1.3$}}

\title[WASP-131: obliquity and stellar rotation]{WASP-131~b with ESPRESSO I: A bloated sub-Saturn on a polar orbit around a differentially rotating solar-type star\thanks{Based on observations made at ESO's VLT (ESO Paranal Observatory, Chile) under ESO programme 106.21EM (PI Cegla) and utilising photometric lightcurves from the Transiting Exoplanet Survey Satellite (TESS), the Next Generation Transit Survey (NGTS) and EulerCam.}}

\author[L. Doyle et al.]{
L. Doyle,$^{1,2}$\thanks{E-mail: lauren.doyle@warwick.ac.uk}
H. M. Cegla,$^{1,2}$\thanks{UKRI Future Leaders Fellow}
D.~R.~Anderson,$^{1,2}$
M. Lendl,$^{3}$
V. Bourrier, $^{3}$
E. Bryant, $^{4}$
J. Vines,$^{5}$
\newauthor
R. Allart, $^{6}$
D. Bayliss,$^{1,2}$
M.~R. Burleigh,$^{7}$
N.~ Buchschacher,$^{3}$ 
S.~L. Casewell,$^{7}$
F. Hawthorn, $^{1,2}$
\newauthor
J.~S. Jenkins,$^{8,9}$
M. Lafarga,$^{1,2}$
M. Moyano,$^{10}$
A. Psaridi,$^{3}$
N. Roguet-Kern, $^{3}$
D.~Sosnowska,$^{3}$
P. Wheatley, $^{1,2}$
\\
$^{1}$Centre for Exoplanets and Habitability, University of Warwick, Coventry, CV4 7AL, UK \\
$^{2}$Department of Physics, University of Warwick, Coventry, CV4 7AL, UK\\
$^{3}$Observatoire Astronomique de l'Universit\'e de Gen\`eve, Chemin Pegasi 51, CH-1290 Versoix, Switzerland \\
$^{4}$Mullard Space Science Laboratory, University College London, Holmbury
St Mary, Dorking, Surrey RH5 6NT, UK \\
$^{5}$ Departamento de Astronom\'ia, Universidad de Chile, Casilla 36-D, 7591245, Santiago, Chile\\
$^{6}$ D\'epartement de Physique, Institut Trottier de Recherche sur les Exoplan\`etes, Universit\'e de Montr\'eal, Montr\'eal, Qu\'ebec, H3T 1J4, Canada \\
$^{7}$ School of Physics and Astronomy, University of Leicester, Leicester LE1 7RH, UK\\
$^{8}$Instituto de Estudios Astrof\'isicos, Facultad de Ingenier\'ia y Ciencias, Universidad Diego Portales, Av. Ej\'ercito 441, Santiago, Chile\\
$^{9}$Centro de Astrof\'isica y Tecnolog\'ias Afines (CATA), Casilla 36-D, Santiago, Chile\\
$^{10}$Instituto de Astronom\'ia, Universidad Cat\'olica del Norte, Angamos 0610, 1270709, Antofagasta, Chile\\
}

\date{Accepted 2023 April 22. Received 2023 March 30; in original form 2023 February 21}

\pubyear{2023}

\begin{document}
\label{firstpage}
\pagerange{\pageref{firstpage}--\pageref{lastpage}}
\maketitle

\begin{abstract}
In this paper, we present observations of two high-resolution transit datasets obtained with ESPRESSO of the bloated sub-Saturn planet WASP-131~b. We have simultaneous photometric observations with NGTS and EulerCam. In addition, we utilised photometric lightcurves from {\tess}, WASP, EulerCam and TRAPPIST of multiple transits to fit for the planetary parameters and update the ephemeris. We spatially resolve the stellar surface of WASP-131 utilising the Reloaded Rossiter McLaughlin technique to search for centre-to-limb convective variations, stellar differential rotation, and to determine the star-planet obliquity for the first time. We find WASP-131 is misaligned on a nearly retrograde orbit with a projected obliquity of $\lambda = 162.4\substack{+1.3 \\ -1.2}^{\circ}$. In addition, we determined a stellar differential rotation shear of $\alpha = 0.61 \pm 0.06$ and disentangled the stellar inclination ($i_* = 40.9\substack{+13.3 \\ -8.5}^{\circ}$) from the projected rotational velocity, resulting in an equatorial velocity of $v_{\rm{eq}} = 7.7\substack{+1.5 \\ -1.3}$~km s$^{-1}$. In turn, we determined the true 3D obliquity of $\psi = 123.7\substack{+12.8 \\ -8.0}^{\circ}$, meaning the planet is on a perpendicular/polar orbit. Therefore, we explored possible mechanisms for the planetary system's formation and evolution. Finally, we searched for centre-to-limb convective variations where there was a null detection, indicating that centre-to-limb convective variations are not prominent in this star or are hidden within red noise.  
\end{abstract}




\section{Introduction}
To understand the conditions and habitats of exoplanets we need to fully understand their stellar hosts. Stellar activity and surface phenomena (such as flares, spots, granulation and faculae/plage) can cause biases in the calculations of planetary properties and can mask/mimic planets in observations. For example, \citet{saar1997activity} and \citet{zhao2022expres} show stellar surface phenomena can alter observed stellar absorption line profiles which may be mistaken for Doppler shifts that can mask and mimic the Doppler motion of a planetary companion. Furthermore, \citet{oshagh2016can} show how occulted starspots affect the Rossiter McLaughlin waveform causing inaccuracies on derived planetary properties. \citet{cegla2016modeling} explores the impact of differential rotation on the projected star-planet obliquity and \citet{cegla2016rossiter} show that if centre-to-limb convective velocity variations are ignored they can bias our measurements of planetary system geometries, which in turn skews our understanding of planetary formation and evolution. 

When a planet transits a host star, a portion of the starlight is blocked in the line-of-sight and a distortion of the velocities is observed, known as the Rossiter-McLaughlin (RM) effect (see \citet{rossiter1924detection, mclaughlin1924some} for original studies and \citet{queloz2000detection} for the first exoplanet case). The Reloaded RM (RRM) technique isolates the blocked starlight behind the planet to spatially resolve the stellar spectrum \citep{cegla2016rossiter}. The isolated starlight from the RRM can be used to derive the projected obliquity, $\lambda$, (i.e. the sky-projected angle between the stellar spin axis and planetary orbital plane). If the planet occults multiple latitudes, we can determine the stellar inclination (by disentangling it from the projected rotational velocity, $v \sin i_*$). Alternatively, we can also use the stellar rotation period (P$_{\rm{rot}}$) in combination with  $v \sin i_*$ to determine the stellar inclination $i_{*}$. This then  helps to measure $\lambda$ and determine the 3D obliquity, $\psi$ (i.e. the angle between the stellar equator and planetary plane). When considering planetary migration/evolution, $\psi$ is of great importance as it avoids introducing biases from only knowing $\lambda$ \citep[see][]{albrecht2021preponderance, albrecht2022stellar}. 

If we disentangle the stellar inclination from $v \sin i_*$, then we can probe the stellar differential rotation (DR) of the star using the RRM method. Dynamo processes are largely responsible for the generation of magnetic fields where DR plays a key role \citep[e.g.][]{kitchatinov2011differential, karak2020stellar}. Overall, understanding DR across various spectral types is not just important for exoplanet characterisation, but for magnetic activity as a whole. There are several techniques which can be used to detect DR. For example, Doppler Imaging \citep{vogt1983doppler} can be used to estimate the location of spots on the stellar surface through their effect on spectral line profiles. This is only sufficient for stars with high rotation rates \citep[see][]{collier2002stellar} as differential rotation is measured from differences in the rotation periods of individual spots at different latitudes. The Fourier transform (FT) method \citep{reiners2002feasibility} is used to measure the Doppler shift at different latitudes due to rotation which can be inferred from the FT of the line profiles. In another method, time series photometry is used to measure the total spread of rotation periods resulting from spots at different latitudes \citep{reinhold2013rotation}. This can be done by following the variation of rotation period over time where different spots at different latitudes show close multiple periods which can be used to determine the differential rotational shear. Finally, transiting planets which frequently occult spots at different latitudes can also be used to probe the differential rotational shear of stars \citep[e.g.][]{silva2011time, araujo2021kepler}. However, many of these techniques depend on the star being active (i.e. possessing starspots) which can reduce the sample of targets available but also can introduce degeneracies within their measurements. The RRM technique, allows for the measurement of DR on quiet solar-type stars and can be more direct and precise. 

In addition to modelling for the projected obliquity and DR, we can also use the isolated starlight to account for any centre-to-limb convective velocity variations (CLV) on the stellar surface. Sun-like stars which possess a convective envelope have surfaces covered in granules, bubbles of hot plasma which rise to the surface (blueshift), before cooling and falling back into intergranular lanes (redshift). The net convective velocity shift caused by these granules changes as a function of limb angle (i.e. from the centre to the limb of the star) due to line-of-sight changes and the corrugated surface of the star. Overall, these velocity shifts (or changes in the contrast of the spectra) can impact the RM effect which is used to determine the projected obliquity \citep{cegla2016modeling, bourrier2017refined} and ignoring these effects can bias or skew our understanding of the formation and evolution of planetary systems. 

In this study, we focus on WASP-131~b which is a transiting bloated sub-Saturn planet, discovered by \citet{hellier2016wasp} with M$_{\rm{p}}$ = 0.27 $\pm$ 0.02 M$_{\rm{Jup}}$ and R$_{\rm{p}}$ = 1.22 $\pm$ 0.05~R$_{\rm{Jup}}$. WASP-131 is a G0 main sequence star with V = 10.1 and T$_{\rm{eff}}$ = 5950 $\pm$100~K. It has an inflated radius of R$_*$ = 1.53 $\pm$ 0.05 \Rsun\ and mass M$_*$ =  1.04~$\pm$~0.04 \Msun\ and when placed on its evolutionary track it has an age between 4.5 -- 10~Gyr \citep{hellier2016wasp}. This system was discovered by WASP-South and followed up by CORALIE with a total of 23 RVs \citep[see][for full details]{hellier2016wasp}. In \citet{bohn2020multiplicity}, they detected a relatively faint ($\Delta K$ = 2.8 $\pm$ 0.2), very close in companion at a separation of $\sim$0.19\arcsec ($\sim$38~AU) using imaging from the VLT/SPHERE/IRDIS, where the probability of it being a background object is $<$0.1\%. Therefore, it is likely a gravitationally bound companion with a derived a mass of 0.62 $\pm$ 0.05~\Msun. In a follow up study \citep{southworth2020multiplicity}, used theoretical spectra to propagate the observed K-band light ratios into the optical passbands used to observe WASP-131 and applied a method to correct the velocity amplitudes of the host stars for the presence of contaminating light. In doing this combined with {\tess} data from Sector 11, they fit for the planetary parameters finding  M$_{\rm{p}}$ = 0.27 $\pm$ 0.02 M$_{\rm{Jup}}$ and R$_{\rm{p}}$ = 1.20 $\pm$ 0.06~R$_{\rm{Jup}}$, which are in excellent agreement with \citet{hellier2016wasp}. Overall, for WASP-131 they found the contaminating star is sufficiently faint and makes an insignificant difference to the derived physical stellar and planetary properties. 

In this paper we apply the RRM technique on newly acquired ESPRESSO observations of the WASP-131~b system. We look to characterise stellar DR, centre-to-limb convection-induced variations, and to determine the star-planet obliquity. In \S \ref{sec:obs} we detail all of the photometric and spectroscopic observations used in this study. In \S \ref{stellar_analysis} we obtain updated stellar parameters from a fitting of the spectral energy distribution. \S \ref{sec:trans_analysis} then details the transit and orbital analysis of {\tess} sector 11 and new NGTS and Euler photometric lightcurves where the transit parameters of the system are derived. Finally, we discuss the RRM analysis and results in \S \ref{sec:rrm} followed by the discussion and conclusions in \S \ref{discuss_conclusions}.

\section{Observations}
\label{sec:obs}
We used photometric data from {\tess}, NGTS, WASP, TRAPPIST and EulerCam as well as spectroscopic data from ESPRESSO to analyse the WASP-131 system. In this section, we detail the observations and data reduction where a summary can be found in Table \ref{tab:observations}.

\begin{table}
    \centering
    \caption{Summary of the ESPRESSO, TESS, NGTS, EulerCam, WASP and TRAPPIST data used in this work.}
    {\sl ESPRESSO}
    \begin{tabular}{ccccccc}
    \hline
    \hline
    Run & Night & $N_{\rm{obs}}$ & $t_{\rm{exp}}$ & $\gamma$$^a$ & SNR$^b$ & $\sigma_{\rm{RV}}$$^c$ \\
    & & & (s) & (km s$^{-1}$) & (550nm) & (cms$^{-1}$) \\
    \hline
    A & 08 Mar 2021 & 73 & 130 & $-$19.6906 & 45 & 214 \\
    B & 24 Mar 2021 & 92 & 130 & $-$19.6801 & 49 & 165 \\
    \hline 
    \end{tabular}

    \vspace{5mm}
        {\sl CORALIE}
    \begin{tabular}{ccccc}
    \hline
    \hline
    Date & $N_{\rm{obs}}$ & $t_{\rm{exp}}$ & SNR$^b$ & $\sigma_{\rm{RV}}$$^c$ \\
    & & (s)  & (550nm) & (cms$^{-1}$) \\
    \hline
    Feb 2014 - Mar 2016  & 23 & 1800 & 50  & 610\\
    \hline 
    \end{tabular}
    
    \vspace{5mm}
    {\sl TESS}
    \begin{tabular}{ccccc}
    \hline 
    \hline 
    Sector & Date & $N_{\rm{obs}}$ & $t_{\rm{exp}}$ & $\sigma_{\rm{residual}}$ \\
    & & & (s) & (ppm\,per\,2\,min) \\
    \hline
    11 & 26 Apr -- 20 May 2019 & 13\,887 & 120 & 447 \\
    \hline
    \end{tabular}
    
    \vspace{5mm}
    {\sl NGTS}
    \begin{tabular}{ccccc}
    \hline 
    \hline 
    No. Cameras & Date & $N_{\rm{obs}}$ & $t_{\rm{exp}}$ & $\sigma_{\rm{residual}}$ \\
    & & & (s) & (ppm\,per\,2\,min) \\
    \hline
    5 & 08 Mar 2021 & 8\,436 & 10 & 1127 \\
    5 & 24 Mar 2021 & 10\,813 & 10 & 777 \\
    \hline 
    \end{tabular}

    \vspace{5mm}
    {\sl EulerCam}
    \begin{tabular}{ccccc}
    \hline 
    \hline 
    Wavelength Filter & Date & $N_{\rm{obs}}$ & $t_{\rm{exp}}$ & $\sigma_{\rm{residual}}$ \\
    & & & (s) & (ppm\,per\,2\,min) \\
    \hline
    Gunn $r$ (RG) filter & 22 Apr 2014 & 313 & 47 & 1075 \\
    $I_c$ filter & 02 Mar 2015 & 272 & 62 & 770 \\
    $I_c$ filter & 03 Apr 2015 & 408 & 38 & 1583 \\
    Gunn $r$ (RG) filter & 24 Mar 2021 & 609 & 30 & 611 \\
    \hline 
    \end{tabular}
    
    \vspace{5mm}
    {\sl WASP}
    \begin{tabular}{cccc}
    \hline 
    \hline 
    Date & $N_{\rm{obs}}$ & $t_{\rm{exp}}$ & $\sigma_{\rm{residual}}$ \\
    & & (s) & (ppm\,per\,2\,min) \\
    \hline
    ~~~ 2007 Feb -- 2012 Jun ~~~~ & 23\,328 & 30 & 4329 \\
    \hline 
    \end{tabular}
    
    \vspace{5mm}
    {\sl TRAPPIST}
    \begin{tabular}{ccccc}
    \hline 
    \hline 
    Wavelength Filter & Date & $N_{\rm{obs}}$ & $t_{\rm{exp}}$ & $\sigma_{\rm{residual}}$ \\
    & & & (s) & (ppm\,per\,2\,min) \\
    \hline
    $z$ band & 22 Apr 2014 & 1107 & 10 & 4798 \\
    $z$ band & 19 Apr 2015 & 1200 & 10 & 3846 \\
    $z'$ band & 06 Jun 2015 & 13\,61 & 10 & 5892 \\
    \hline 
    \end{tabular}
    
        \vspace{2mm}
     \begin{flushleft}
    {\bf Notes:} $^a$ The error on the systemic velocity, $\gamma$, is $\sim$0.015~km s$^{-1}$ for each run. $^b$ The SNR per-pixel was computed as the average SNR for order 112 of all observations where 550~nm falls on. $^c$ $\sigma$ stands for the average uncertainty of the disk integrated RVs. Run B was affected by an atmospheric dispersion issue discussed in \S \ref{spec:data}. 
    \end{flushleft}
    \label{tab:observations}
\end{table}

\subsection{Photometric Data}
For WASP-131~b, we utilised four {\tess}, two NGTS, four Euler, 23 WASP and three TRAPPIST photometric transits to determine the transit planet properties. 

\subsubsection{TESS Photometry}
WASP-131 was observed by the Transiting Exoplanet Survey Satellite \citep[{\tess}:][]{ricker2014tess} in 2-min cadence capturing a total of four transits during Sector 11 between the 22 April 2019 and 20 May 2019, see Figure \ref{fig:all_lcs}. We accessed the 2-min lightcurves produced by the TESS Science Processing Operations Centre (SPOC) pipeline \citep{jenkins2016spoc} and used the \textsc{PDCSAP\_FLUX} time series for our analysis. A visual inspection of the lightcurve showed no evidence for a stellar rotation period, however, we also ran a Lomb-Scargle analysis which yielded no significant periodic signals. There is some low level variability which could be either astrophysical or instrumental. Therefore, we removed this variability before fitting the transits by flattening the lightcurve using a spline fit, masking out the in-transit data points to avoid the spline affecting the transit shapes and applying to all data.

\subsubsection{NGTS Photometry}
The Next Generation Transit Survey \citep[NGTS: ][]{wheatley2018ngts} is a photometric facility consisting of twelve independently operated robotic telescopes at ESO's Paranal Observatory in Chile. This is the same location as ESPRESSO, therefore, both instruments experience the same weather conditions. Each NGTS telescope has a 20\,cm diameter aperture and observes using a custom NGTS filter (520 -- 890\,nm).

We observed WASP-131 using NGTS simultaneously with the ESPRESSO transit observations on the nights of 8th March 2021 and 24th March 2021 (see Figure \ref{fig:all_lcs}). For both NGTS observations, we utilised five NGTS telescopes simultaneously to observe the transit event in the multi-telescope observing mode described in \citet{smith2020shallow} and \citet{bryant20multicam}. For both nights, an exposure time of $10\,$seconds was used and the star was observed at airmass $<$ 2. On the night of 8th March 2021 a total of 8436 images were taken across the five telescopes, and 10813 were taken on the night of 24th March 2021. 

The NGTS images were reduced using a version of the standard NGTS photometry pipeline, described in \citet{wheatley2018ngts}, which has been adapted for targeted single star observations. In short, standard aperture photometry routines are performed on the images using the SEP Python library \citep{bertin1996sextractor, barbary16sep}. For the WASP-131 observations presented in this work, we used circular apertures with a radius of five~pixels (25\,\arcsec). During this reduction comparison stars which were similar in magnitude and CCD position to the target star were automatically identified using the \textit{Gaia} DR2 \citep{GAIA, GAIA_DR2} and parameters found in the {\tess} input catalogue \citep[v8;][]{stassun2019ticv8}. Each comparison star selected was also checked to ensure it is isolated from other stars.

\subsubsection{Euler Photometry}
EulerCam \citep{lendl2012wasp} is an $e$2$v$ 4k $\times$ 4k back-illuminated deep-depletion silicon CCD detector installed in 2010 at the Cassegrain focus of the 1.2-m Euler-Swiss telescope located at La Silla in Chile. The total field of view is 15.68 $\times$ 15.73~arcmin, with a resolution of 0.23~arcsec per pixel. WASP-131~b was observed four times by EulerCam on the 22nd April 2014 (\emph{Gunn} $r'$ filter), 2nd March 2015 ($I_c$ filter), 3rd April 2015 ($I_c$ filter) and simultaneously with ESPRESSO on 24th March 2020 (\emph{Gunn} $r'$ filter). All four lightcurves can be seen in Figure \ref{fig:all_lcs} along with the transit fitting. 

\subsubsection{TRAPPIST Photometry}
The TRAnsiting Planets and PlanetesImals Small Telescope \citep[TRAPPIST:][]{gillon2011detection, gillon2011wasp} is a 60-cm robotic telescope at La Silla in Chile. Its thermoelectrically-cooled camera is equipped with a 2k $\times$ 2k Fairchild 3041 CCD. This provides a field of view of 22 $\times$ 22~arcmin with a 0.65~arcsec scale pixel. TRAPPIST observed three transits of WASP-131~b on 22nd April 2014 ($z$ band), 19th April 2015 ($z$ band) and 6th June 2015 ($z'$ band). All lightcurves can be seen in Figure \ref{fig:all_lcs} along with the transit fitting.  

\subsubsection{WASP Photometry}
\begin{figure}
    \centering
    \includegraphics[width=0.48\textwidth]{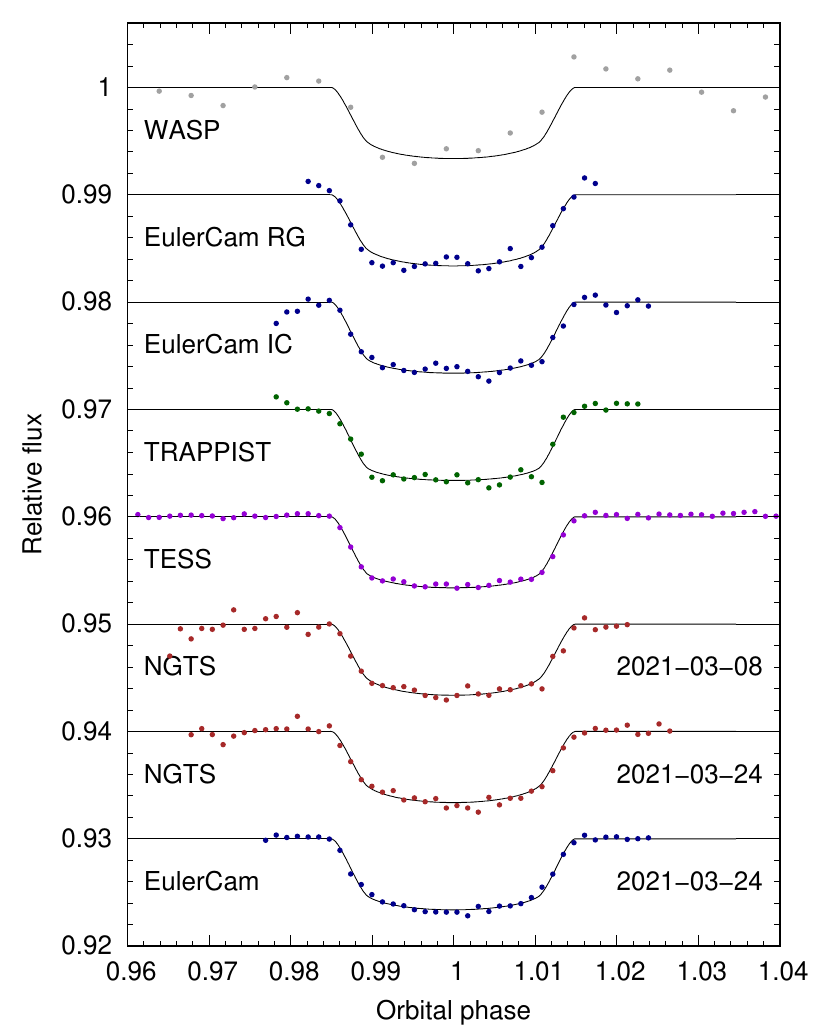}
    \caption{All lightcurves of WASP-131 from WASP, EulerCam, TRAPPIST, {\tess} and NGTS together with the fitted transit model using properties from Table \ref{planetary_properties}. All data is binned to 10 minutes with the exception of the WASP data which is binned to 30 minutes. The WASP, EulerCam IC, TRAPPIST and TESS lightcurves all contain multiple phase folded transits, whereas the NGTS EulerCam RG and EulerCam are single transits. The bottom three plots are the NGTS and EulerCam data observations which were simultaneous to the ESPRESSO transits. }
    \label{fig:all_lcs}
\end{figure}

The WASP survey was operated from two sites with one in each hemisphere. The data here was collected by WASP-South based in the Sutherland Station of the South African Astronomical Observatory (SAAO). Each site consists of eight commercial 11-cm ($f =$200~mm) Canon lenses backed by 2k $\times$ 2k Peltier-cooled CCDs on a single mount. This provides a field of 7.8$^{\circ}$ $\times$ 7.8$^{\circ}$ with a typical cadence of 10 minutes. Full details of the WASP survey and the photometric reduction can be found in \citet{pollacco2006wasp} and \citet{collier2007efficient}. WASP-131 was observed between Feb 2007 and June 2012 from WASP-South obtaining a total of 23\,328 data points. The phase folded and binned WASP lightcurve can be seen in Figure \ref{fig:all_lcs}.

\begin{table}
    \centering
    \caption{System parameters for WASP-131}
    \begin{tabular}{l|c}
    \hline 
    Parameter (unit) & Value \\
    \hline
    {\bf Stellar parameters from {\sc ariadne}} & \\
    \hline
    $T_{\rm eff}$ (K) & $5990 \pm 50$\\
    ${\rm [Fe/H]}$ & $-0.20 \pm 0.07$\\
    $\log g_*$ & $3.89 \pm 0.08$\\ 
    $A_V$ & $0.05 \pm 0.03$\\
    $M_{\rm *}$ ($M_{\rm \odot}$) & $1.06 \pm 0.06$\\
    $R_{\rm *}$ ($R_{\rm \odot}$) & $1.68 \pm 0.02$\\
    Age (Gyr) & $7 \pm 1$\\
    \hline 
    {\bf MCMC proposal parameters} & \\
    \hline
    $T_{\rm{0}}$ (BJD) & \Ntc \\
    $P_{\rm{orb}}$ (days) & \Nperiod \\
    $T_{\rm{dur}}$ (hours) & \Nduration \\
    $(R_{\rm{p}} / R_{\rm{*}})^2$ & \Nrprssq \\
    $b$ & \Nimpact \\
    K (m s$^{-1}$) & \Nreflex \\
    $\gamma$ (m s$^{-1}$) & \Ngamma \\
    $e$ & 0 (adopted) \\
    \hline
    {\bf MCMC derived parameters} & \\
    \hline
    $a/R_*$ & \Nars \\
    $i_{\rm{p}}$ ($^{\circ}$) & \Ninc \\
    $M_{\rm *}$ ($M_{\rm \odot}$) & \Nmstar\\
    $R_{\rm *}$ (\Rsun) & \Nrstar \\
    $\log g_*$ (cgs) & \Nloggstar \\
    $\rho_*$ ($\rho_{\odot}$) & \Nrhostar \\
    $M_{\rm p}$ ($M_{\rm \odot}$) & \Nmplanet\\
    $R_{\rm p}$ (\Rsun) & \Nrplanet \\
    $\log g_p$ (cgs) & \Nloggplanet \\
    $\rho_p$ ($\rho_{\rm Jup}$) & \Nrhoplanet \\
    a (AU) & \Naau \\
    \hline
    {\it Limb-darkening coefficients:} & \\
    \hline
    WASP/Euler (RG)/NGTS& c1=0.607, c2=-0.115, \\
    & c3=0.562, c4=$-0.318$ \\
    Euler ($I_c$)/TESS& c1=0.683, c2=$-0.380$, \\
    & c3=0.723, c4=$-0.363$ \\
    TRAPPIST($z$ band)& c1=0.580, c2=$-0.171$, \\
    & c3=0.450, c4=$-0.258$ \\
    $u_1$ & \Nuonets \\
    $u_2$ & \Nutwots \\
    \hline
    \end{tabular}
    \begin{flushleft}
    {\bf Notes:} The limb darkening coefficients $u_1$ and $u_2$ were obtained in the ESPRESSO passband (380 -- 788~nm) by inputting the WASP-131 stellar parameters into the ExoCTK  calculator \citep{matthew_bourque}.
    \end{flushleft}
    \label{planetary_properties}
\end{table}

\subsection{Spectroscopic Data}
\label{spec:data}
Two transits of WASP-131~b were observed on the nights of 8th March 2021 (run A) and 24th March 2021 (run B) using the ESPRESSO \citep{pepe2014espresso, pepe2021espresso} spectrograph (380 - 788~nm) mounted on the Very Large Telescope (VLT) at the ESO Paranal Observatory in Chile (ID: 106.21EM, PI: H.M. Cegla). The ESPRESSO observations were carried out using UT3 for the first night and UT1 for the second under good conditions, with airmass varying between 1.0 -- 2.4'' and 1.0 -- 2.2'' for each run A and run B, respectively, in the high resolution mode (R $\sim$138,000) using 2 $\times$ 1 binning. Exposure times were fixed at 130~s for each night to reach a signal-to-noise (SNR) near 50 at 550~nm (to be photon noise dominated) and to ensure a good temporal cadence, with a 45~s readout time per exposure. Each run respectively covered a duration of 6h 54m and 6h 59m of uninterrupted sequences covering the full transit duration and includes $\sim$1~hr pre- and $\sim$1~hr post- baseline. A summary of the ESPRESSO observations can be found in Table \ref{tab:observations}. 

The spectra were reduced with version 2.2.8 of the ESPRESSO data reduction software\footnote{\href{www.eso.org/sci/software/pipelines/espresso/ espresso-pipe-recipes.html}{www.eso.org/sci/software/pipelines/espresso/ espresso-pipe-recipes.html}} \citep[DRS,][]{pepe2021espresso}, using a F9 binary mask (F9 is the closest to G0 of all spectral types available) to cross-correlate the observed spectra to generate high SNR cross-correlation functions (CCFs) which we used for our analysis. Additionally, the DRS also outputs the relative depth (contrast), full width at half max (FWHM) and radial velocity centroid of each CCF. In run A the SNR can be seen to increase over the duration of the night which correlates with the decreasing airmass as observing conditions improve. In run B the SNR increases sharply at the beginning of the night where it then remains relatively stable throughout the transit observation. Furthermore, the FWHM and contrast remain steady during both runs and are dispersed around the mean. Overall, the average integrated radial velocity uncertainties for run A and run B are 2.14~ms$^{-1}$ and 1.65~ms$^{-1}$, respectively.

During run B (night of 24th March 2021) ESPRESSO was affected by low-level software issues, at the level of communication with the Programmable Logic Controller which did not trigger any error or warning. This resulted in the Atmospheric Dispersion Corrector (ADC) being non-responsive preventing the correction of the atmospheric dispersion, which in turn introduced a wavelength-dependent light loss at the fibre interface. A comparison by ESO of RVs taken with the ADC operating in and out of range (i.e., correcting all atmospheric dispersion or leaving uncorrected atmospheric dispersion) respectively, shows that the latter dataset is affected by an additional scatter on the order of $\sim$1~ms$^{-1}$ (note this is not target specific). Overall, we do not notice a difference between the behaviour of the two nights, therefore, we do not foresee this as an issue on our analysis or results. 

\section{Stellar parameters}
\label{stellar_analysis}
\begin{figure}
	\includegraphics[width=\columnwidth]{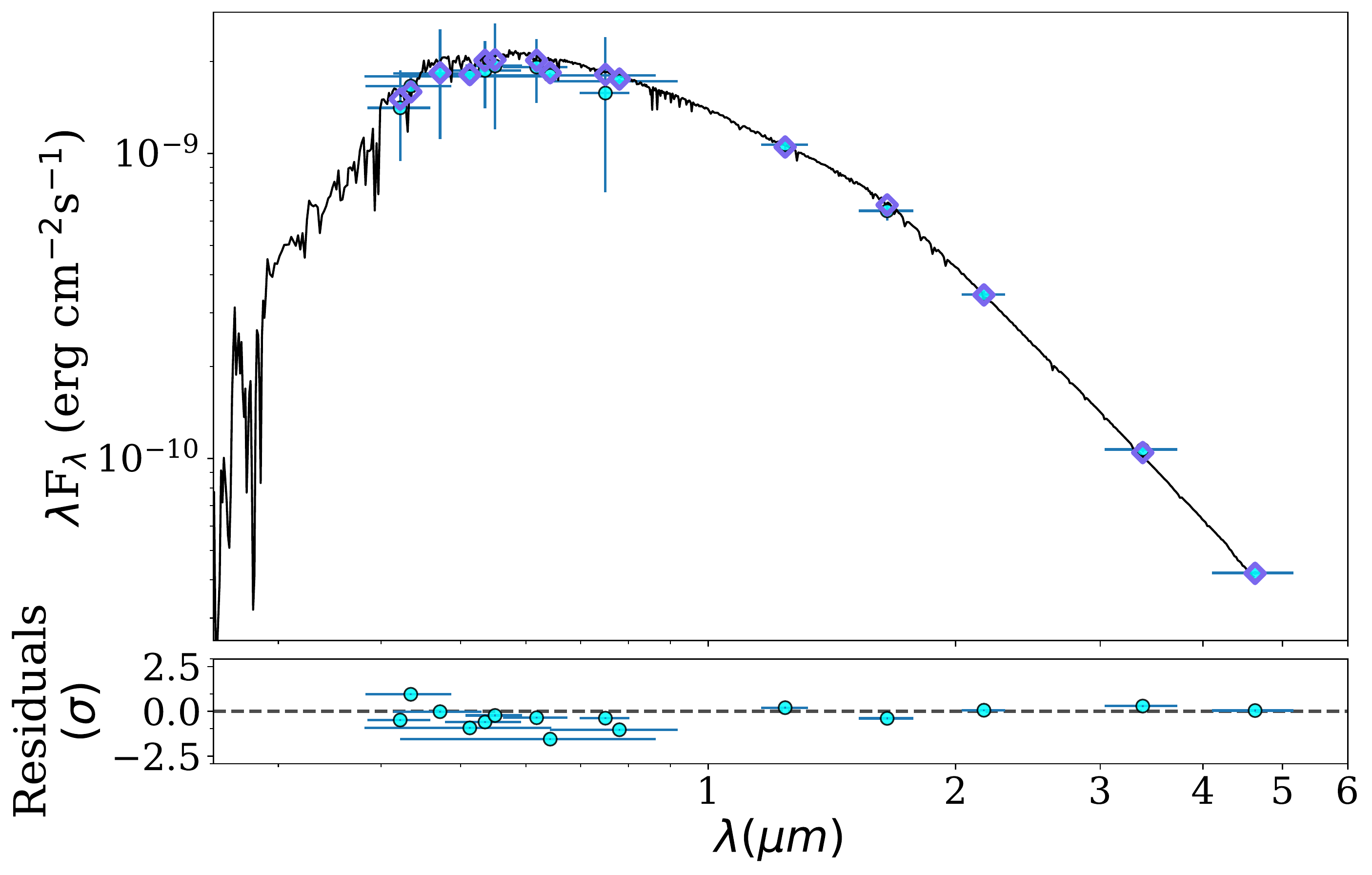}
    \caption{{\it Top:} The best-fitting SED (black line) for WASP-131 to catalog photometry (cyan points). The bandpass widths are indicated by horizontal error bars. The purple diamonds show the synthetic fluxes at the wavelengths of the photometric data. {\it Bottom:} The residuals after subtracting the best-fitting model.}
    \label{fig:SED}
\end{figure}

We used {\sc ariadne}\footnote{\href{https://github.com/jvines/astroARIADNE}{https://github.com/jvines/astroARIADNE}} \citep{ariadne} to obtain stellar parameters by fitting the spectral energy distribution (SED) of WASP-131, as sampled by catalogue photometry, with stellar atmosphere models. We placed Gaussian priors on the stellar parameters, using the values from \citet{hellier2016wasp}, and on stellar distance, using the estimate from Gaia eDR3 \citep{Bailer-Jones2021}. We placed uniform priors on stellar radius (0.5 to 20 R$_\odot$) and on extinction (0 to 0.138; \citealt{Schlegel1998, Schlafly2011}). We applied Bayesian model averaging to the results obtained when using the following four stellar atmosphere models: Phoenix V2 \citep{Husser2013}, BT-Settl \citep{Allard2012}, \citet{Castelli2004}, and \citet{Kurucz}. The SED of WASP-131 in shown in Figure \ref{fig:SED} and the values of key parameters are given in Table \ref{planetary_properties}.

\section{Transit and orbital analysis}
\label{sec:trans_analysis}
To fit the transit lightcurves we follow the method of \citet{hellier2016wasp} who originally determined the properties of the WASP-131 system. Since their study, there has been new {\tess} data released along with the simultaneous lightcurves to ESPRESSO obtained by NGTS and EulerCam. \citet{southworth2020multiplicity} used the {\tess} data to correct for contaminating light from the companion star and found it to be sufficiently faint to make little difference to measurements of WASP-131~b's physical properties. Therefore, we combined the new lightcurves from {\tess}, NGTS and EulerCam along with the original datasets to obtain updated planetary parameters using the method of \citet{hellier2016wasp}, including the most precise ephemeris possible. 

The WASP, EulerCam, {\tess}, NGTS and TRAPPIST photometry were combined with the CORALIE radial velocity measurements from \citet{hellier2016wasp} in a simultaneous Markov Chain Monte Carlo (MCMC) analysis to determine the planetary parameters. Full details of this method can be found in \citet{cameron2007wasp}. We interpolated the tabulations of \citet{claret2000new} and \citet{claret2004} to obtain coefficients for the four-parameter, non-linear limb-darkening law (Table \ref{planetary_properties}). We also determined quadratic limb darkening parameters in the ESPRESSO passband (380 -- 788~nm) by inputting the WASP-131 stellar parameters into the ExoCTK calculator \citep{matthew_bourque} using Top Hat which assumes 100 per cent throughput at all wavelengths. These determined limb darkening parameters within the ESPRESSO bandpass are used for the model lightcurve which scales the CCF before the direct subtraction between in-transit and out-of-transit observations. 

The fitted transit parameters were $T_0$, $P_{\rm{orb}}$, $R_{\rm{p}}^2 / R_{\rm{*}}^2$, $T_{\rm{dur}}$, $b = a \cos i_p / R_*$, where $T_0$ is the epoch of mid-transit, $P_{\rm{orb}}$ is the orbital period, $R_{\rm{p}}^2 / R_{\rm{*}}^2$ is the planet-to-star radius ratio squared, $T_{\rm{dur}}$ is the total transit duration (from first to fourth contact), $b$ is the impact parameter in the case of a circular orbit, $a$ is the semi-major axis and $i_p$ is the orbital inclination. The eccentric Keplerian orbit was parameterised by the stellar reflex velocity semi-amplitude $K_1$, the systemic velocity $\gamma$, and $\sqrt{e} \cos \omega$ and $\sqrt{e} \sin \omega$, where $e$ is orbital eccentricity and $\omega$ is the argument of periastron. 

We tested whether the RV data are best described by a circular or eccentric orbital model. 
The log-evidence is larger for the circular model than the eccentric model ($\Delta \ln Z \sim 3$; odds ratio $\sim$ 20), with the eccentric model favouring a small eccentricity consistent with zero ($e = 0.04^{+0.04}_{-0.03}$). Therefore, we followed \citet{hellier2016wasp} and assumed a circular orbit. We placed a Gaussian prior on stellar mass using the value obtained from the {\sc ariadne} analysis. The fitted and derived parameters, along with their 1$\sigma$ errors, are listed in Table \ref{planetary_properties}.

\section{Reloaded Rossiter McLaughlin}
\label{sec:rrm}
We utilised the RRM technique to isolate the starlight of WASP-131 behind the planet during its transit. A detailed comprehensive description of the technique can be found in \citet{cegla2016rossiter} and \citet{doyle2022WASP-166}. From here on we will use the term local CCF (\locccf) to refer to the occulted light emitted behind the planet and the term disk-integrated CCF (\DIccf) to refer to the light emitted by the entire stellar disk. 

\begin{figure}
    \centering
    \includegraphics[width = 0.47\textwidth]{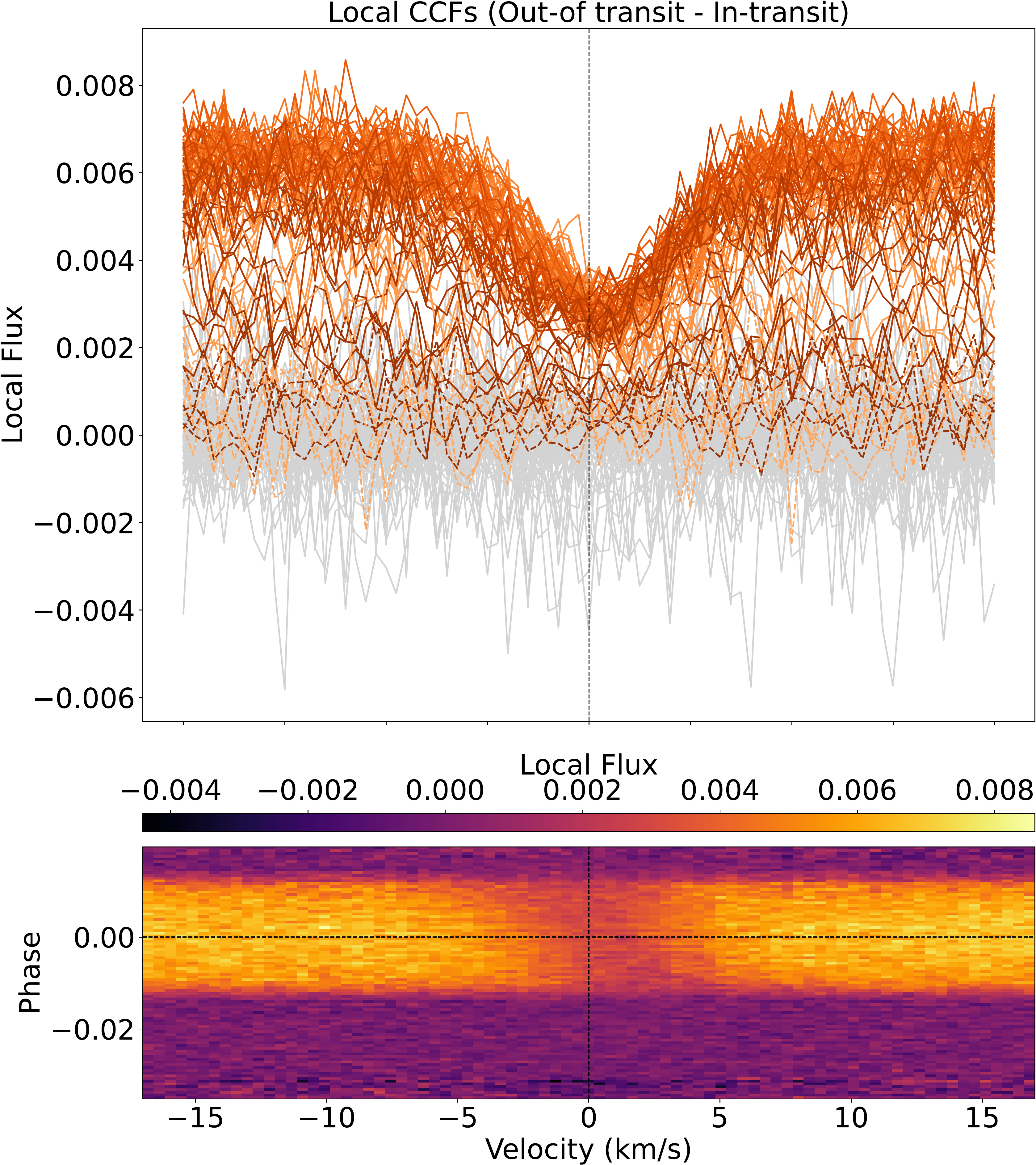}
    \caption{{\it Top:} The local CCFs (out-of-transit – in-transit) in the stellar rest frame of the star behind WASP-131 b. The light grey are the out-of-transit observations and the orange lines are the in-transit observations. The changing gradient of the orange lines represents the changing centroid position where the darker orange is more redshifted. Dashed orange lines are observations which have a stellar disk position $\langle\mu\rangle <$ 0.20 and are not used in the analysis. 
 {\it Bottom:} A top view of the top plot showing a map of the local CCFs colour-coded by the local flux. A dotted line at phase zero and 0~km s$^{-1}$ in both plots is included to guide the eye.   }
    \label{local_ccfs}
\end{figure}

To begin, the ESPRESSO \DIccfs are shifted and re-binned in velocity space to correct for the Keplerian motions of the star induced by WASP-131~b (using the orbital properties in Table \ref{planetary_properties}). A single master-out \DIccf was created for each run by summing all out-of-transit \DIccfs and normalising the continuum to unity which was then fitted by a Gaussian profile to determine the systemic velocity, $\gamma$, see Table \ref{tab:observations}. All \DIccfs were shifted to the stellar rest frame by subtracting $\gamma$ for each corresponding night. Each \DIccf was normalised by their individual continuum value and scaled using a quadratic limb darkened transit model from the fitted parameters in Table \ref{planetary_properties}. Finally, the \locccfs were obtained by subtracting the now scaled in-transit \DIccfs from the master-out \DIccf for each night, see Figure \ref{local_ccfs}. 

\begin{figure*}
    \centering
    \includegraphics[width = 0.97\textwidth]{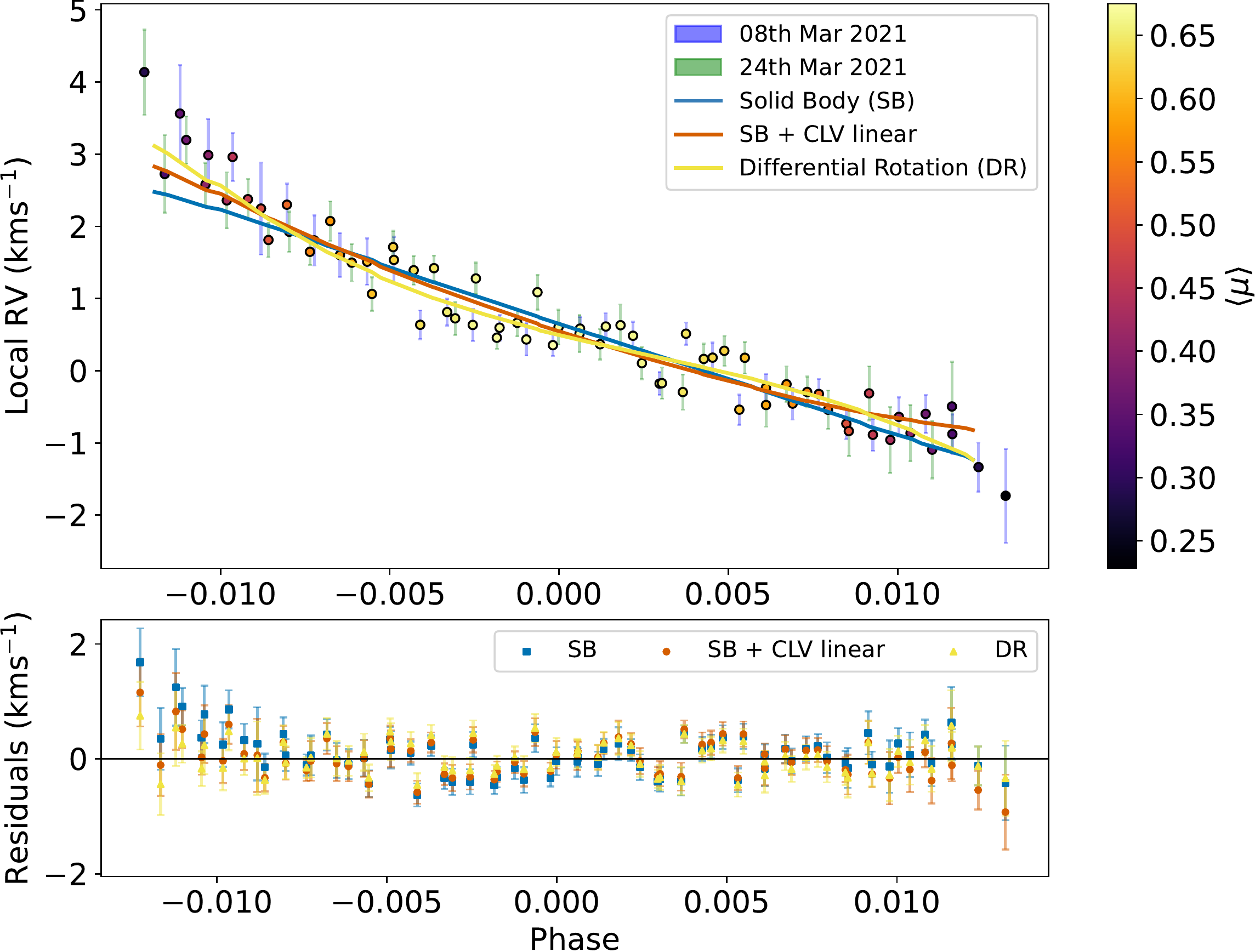}
    \caption{The top panel shows the local RVs determined from the local CCFs of the regions occulted by the planet as a function of phase. The data points are colour coded by the stellar disk position behind the planet in units of brightness weighted $\langle\mu\rangle$ (where $\mu = \cos\theta$). The best-fit model for differential rotation (DR: red line) is shown, along with the solid body (blue) and SB plus centre-to-limb linear (green) models. The bottom panel shows the residuals (local RVs - model) for all models with colours corresponding to the top panel model lines, with a horizontal line at 0 to guide the eye.}
    \label{local_RVs}
\end{figure*}

The stellar velocity of the occulted starlight was determined by fitting Gaussian profiles using {\tt curve\_fit} from the {\tt Python Scipy} package \citep{2020SciPy-NMeth} to each of the \locccfs. Overall, there are a total of four Gaussian parameters in our fit including the offset (i.e. continuum), amplitude, centroid, and FWHM. The flux errors assigned to each \locccf were propagated from the errors on each \DIccf as determined from the version 2.2.8 of the ESPRESSO DRS and included in our Gaussian fit. Figure \ref{local_RVs} shows the resulting local RVs of the planet occulted starlight, plotted as a function of both phase and stellar disk position behind the planet. We removed CCFs with limb angle $\mu < 0.20$ (i.e. the distance from the centre to the limb of the star where $\mu = \cos\theta$) from our analysis, resulting in 12 CCFs being removed from the observations. This was due to profiles close to the limb being very noisy and when comparing the depth to the noise the signal was not significant to enable a Gaussian fit, see Figure \ref{local_ccfs} where they are shown as dashed lines.

We fitted the local RVs in Figure \ref{local_RVs} using the model {\bf and coordinate system} described in \citet{cegla2016rossiter}. This fitting depends on the position of the transiting planet centre with respect to the stellar disk, projected obliquity ($\lambda$), stellar inclination ($i_*$), the equatorial rotational velocity ($v_{eq}$), the differential rotational shear ($\alpha$, ratio between the equatorial and polar stellar rotational velocities), quadratic stellar limb darkening (u$_1$ and u$_2$), and centre-to-limb convective variations of the star ($v_{\rm{conv}}$). For WASP-131, we fitted for different scenarios depending on whether or not we account for DR and centre-to-limb convective variations. 

\subsection{Results}
\label{sec:results}
In Figure \ref{local_RVs}, the measured local RVs decrease with orbital phase from approximately +4~km s$^{-1}$ to -2~km s$^{-1}$ as the planet transits the stellar disk. There is a lack of symmetry within the velocities, where the planet spends more time crossing the red-shifted regions than the blue-shifted regions. Overall, this suggests the WASP-131 system is likely misaligned. In this section we discuss the various stellar rotation scenarios we fitted to the local RVs along with any potential detections of centre-to-limb convective variations. 

\begin{table*}
    \centering
        \caption{MCMC observational results for WASP-131 and the derived 3D spin-orbit obliquity.}
    \resizebox{1.0\textwidth}{!}{    
    \begin{tabular}{lcccccccccccccc}
    \hline
        Model & No. of Model & $v_{\rm{eq}}$ & $i_{\rm{*}}$ & $\alpha$ & $\lambda$    & $\sigma$     &  $c_1$        & $c_2$        & $c_3$        & BIC & \chisq &  $\chi_{\nu}^{2}$ &  $\psi$  \\
              & Parameters   & (km s$^{-1}$)  & ($^{\circ}$) &          & ($^{\circ}$) & (km s$^{-1}$) &(km s$^{-1}$) & (km s$^{-1}$)&   (km s$^{-1}$)&     &        &                   &($^{\circ}$) \\
        \hline
        {\bf Un-binned observations} &&&&&&&&&&\\
        \hline
        SB & 2 &  $3.03 \pm 0.09$ & 90.0 & 0.0 & $163.2 \pm 0.8$ & -- & -- & -- & -- & 142 & 133 & 1.85 &-- \\
        SB + $\sigma$ & 3 &  $3.14 \pm 0.13$ & 90.0 & 0.0 & $162.9 \pm 1.1$ & $0.23 \pm 0.04$ & -- & -- & -- & 80.6 & 67.7 & 0.94 & -- \\
        {\bf SB + CLV1} & {\bf 3} & {\bf 3.05 $\pm$ 0.09} & {\bf 90.0} & {\bf 0.0} & {\bf 167.8 $\pm$ 1.2} & {\bf --} & {\bf -1.5 $\pm$ 0.3} & {\bf --} & {\bf --} & {\bf 123} & {\bf 110} & {\bf 1.5} & {\bf --} \\
        SB + CLV2 & 4 & $3.03 \pm 0.11$ & 90.0 & 0.0 & $171.4\substack{+4.6 \\ -4.8}$ & -- & $0.75\substack{+2.7 \\ -2.6}$ & $-2.1\substack{+2.8 \\ -2.6}$ & -- & 275 & 259 & 3.6 & -- \\
        SB + CLV3 & 5 & $3.13\substack{+0.29 \\ -0.16}$ & 90.0 & 0.0 & $163.3\substack{+10.7 \\ -12.5}$ & -- & $10.2\substack{+15.2 \\ 12.9}$ & $-21.4\substack{+25.3 \\ -31.1}$ & $12.7\substack{+20.6 \\ -16.3}$ & 130 & 109 & 1.5 & -- \\
        {\bf DR} & {\bf 4} & {\bf 7.7$\substack{{\bf +1.5} \\ {\bf -1.3}}$} & {\bf 40.9$\substack{{\bf +13.3} \\ {\bf -8.5}}$} & {\bf 0.61 $\pm$ 0.06} & {\bf 162.4$\substack{{\bf +1.3} \\ {\bf -1.2}}$} & {\bf --} & {\bf --} & {\bf --} & {\bf --} & {\bf 111} & {\bf 94} & {\bf 1.3} & {\bf 123.7$\substack{{\bf +12.8} \\ {\bf -8.0}}$} \\
        DR + $\sigma$ & 5 & $10.7\substack{+11.1 \\ -4.1}$ & $26.9\substack{+25.5 \\ -14.7}$ & $0.68\substack{+0.16 \\ -0.11}$ & $161.5\substack{+1.9 \\ -1.6}$ & $0.17 \pm 0.04$ & -- & -- & -- & 79.9 & 58.5 & 0.81 & $110.4\substack{+24.4 \\ -13.8}$ \\
        DR + CLV1 & 5 & $7.6\substack{+1.5 \\ -1.3}$ & $41.4\substack{+14.8 \\ -9.5}$ & $0.62\substack{+0.15 \\ -0.13}$ & $162.4\substack{+3.1 \\ -5.7}$ & -- & $0.02\substack{+1 \\ -0.8}$ & -- & -- & 116 & 95 &  1.3 & $124.2\substack{+14.8 \\ -9.8}$ \\
        DR + CLV2 & 6 & $7.8 \pm 1.5$ & $38.8\substack{+17 \\ -8}$ & $0.61 \pm 0.14$ & $160.9\substack{+8.5 \\ -9.8}$ & -- & $-0.40 \pm 2.4$ & $0.55\substack{+2.5 \\ -2.8}$ & -- & 120 & 95 & 1.3 & $121.4\substack{+18.2 \\ -9.5}$ \\
        DR + CLV3 & 7 & $7.9\substack{+1.4 \\ -1.5}$ & $46.21\substack{+23 \\ -13}$ & $0.64 \pm 0.19$ & $165.47\substack{+10 \\ -12}$ & -- & $-10.2\substack{+11.6 \\ -10.3}$ & $18.4\substack{+18.1 \\ -21.6}$ & $-10.8\substack{+12.9 \\ -10.2}$ & 126 & 97 & 1.3 & $129.4\substack{+24.5 \\ -14.8}$ \\
        \hline
        {\bf Binned 12 minute observations} &&&&&&&&&&\\
        \hline
        SB & 2 &  $3.01 \pm 0.1$ & 90.0 & 0.0 & $162.9\substack{+0.8 \\ -0.9}$ & -- & -- & -- & -- & 68.2 & 61.2 & 1.9 &-- \\
        SB + $\sigma$ & 3 &  $3.08 \pm 0.14$ & 90.0 & 0.0 & $162.7\substack{+1.2 \\ -1.3}$ & $0.17 \pm 0.05$ & -- & -- & -- & 41.7 & 31.3 & 0.97 & -- \\
        SB + CLV1 & 3 & $3.02 \pm 0.1$ & 90.0 & 0.0 & $166.6 \pm 1.2$ & -- & $-1.3 \pm 0.3$ & -- & -- & 56.2 & 45.8 & 1.4 & -- \\
        SB + CLV2 & 4 & $3.02 \pm 0.12$ & 90.0 & 0.0 & $166.9\substack{+5.3 \\ -5.1}$ & -- & $-1.1\substack{+3.2 \\ -3.3}$ & $-0.16\substack{+3.1 \\ -3.0}$ & -- & 102 & 87.8 & 2.7 & -- \\
        SB + CLV3 & 5 & $3.12\substack{+0.36 \\ -0.18}$ & 90.0 & 0.0 & $160.8\substack{+12.2 \\ -13.6}$ & -- & $6.3\substack{+18.6 \\ -15.6}$ & $-15.1\substack{+30.5 \\ -37.6}$ & $9.8\substack{+24.7 \\ -19.4}$ & 63.4 & 46.0 & 1.4 & -- \\
        DR & 4 & $7.1\substack{+1.8 \\ -1.5}$ & $40.9\substack{+17.4 \\ -10.3}$ & $0.54\substack{+0.07 \\ -0.08}$ & $162.1\substack{+1.3 \\ -1.2}$ & -- & -- & -- & -- & 50.5 & 36.6 & 1.1 & $123.7\substack{+16.5 \\ -9.7}$ \\
        DR + $\sigma$ & 5 & $9.8\substack{+13 \\ -3.9}$ & $27.6\substack{+28.3 \\ -16.6}$ & $0.62\substack{+0.16 \\ -0.12}$ & $161.5\substack{+1.7 \\ -1.6}$ & $0.10 \pm 0.05$ & -- & -- & -- & 43.7 & 26.4 & 0.82 & $111.1\substack{+26.8 \\ -15.6}$ \\
        DR + CLV1 & 5 & $6.7\substack{+2.0 \\ -1.8}$ & 45.9$\substack{+28.2 \\ -14.9}$ & $0.54\substack{+0.2 \\ -0.3}$ & $162.7\substack{+3.3 \\ -5.8}$ & -- & $-0.13\substack{+1.4 \\ -0.96}$ & -- & -- & 54.1 & 36.8 &  1.2 & $128.5\substack{+26.6 \\ -14.9}$ \\
        DR + CLV2 & 6 & $7.2\substack{+1.9 \\ -2.2}$ & $39.6\substack{+31.2 \\ -9.5}$ & $0.57\substack{+0.18 \\ -0.26}$ & $156.7\substack{+9.5 \\ -10.7}$ & -- & $-1.8 \pm 2.9$ & $2.1\substack{+2.7 \\ -3.1}$ & -- & 88.1 & 67.3 & 2.1 & $121.0\substack{+31.3 \\ -11.1}$ \\
        DR + CLV3 & 7 & $7.5\substack{+1.5 \\ -1.6}$ & $44.5\substack{+21.4 \\ -12.5}$ & $0.63\substack{+0.21 \\ -0.20}$ & $162.3\substack{+11.9 \\ -12.6}$ & -- & $-15.2\substack{+13.4 \\ -12.3}$ & $26.9\substack{+21.9 \\ -24.6}$ & $-15.2\substack{+14.6 \\ -15.5}$ & 66.0 & 41.7 & 1.3 & $127.0\substack{+23.4 \\ -14.5}$ \\
        \hline
    \end{tabular}}
    \label{mcmc_results}
    \vspace{2mm}
     \begin{flushleft}
   {\bf Notes:} For all SB models $i_*$ and $\alpha$ are fixed under the assumption of rigid body rotation and the $v_{\rm{eq}}$ column corresponds to $v_{\rm{eq}}\sin i_*$. For these models we are unable to determine the 3D obliquity, $\psi$. The BIC of each model was calculated using \chisq and the reduced chi-squared ($\chi_{\nu}^{2}$) has been added as well to allow for comparisons between the binned and un-binned datasets. For clarity, CLV1, CLV2 and CLV3 correspond to centre-to-limb linear, quadratic and cubic respectively. The best fit model for SB and DR have been highlighted in bold. Corner plots for the un-binned observation MCMC runs for SB, SB plus linear CLV and DR are in an appendix and the remaining are available as supplementary material online.    
    \end{flushleft}
    \label{mcmc_fits}
\end{table*}

\subsection{Solid Body Stellar Rotation}
Firstly, we fit a solid body (SB) stellar rotation model as it is the simplest of models with the least free parameters. The two free parameters for this model are $\lambda$, and $v_{\rm{eq}}\sin i_*$. By fitting both ESPRESSO runs together we find, $v_{\rm{eq}}\sin i_* = 3.03 \pm 0.09$~km s$^{-1}$ and $\lambda = 163.2 \pm 0.8^{\circ}$. The projected stellar velocity is consistent with that of \citet{hellier2016wasp} where they find $v_{\rm{eq}}\sin i_* = 3.00 \pm 0.9$~km s$^{-1}$ and in addition to this, we find the projected obliquity is largely misaligned. 

We are also interested in how the net convective blueshift (CB) varies across the stellar disk. To model these centre-to-limb convective variations (CLV) we fit the local RVs for both SB and CLV at the same time. Since we do not know the shape of the trend of the CLV, we test a linear, quadratic or cubic polynomial as a function of limb angle following the formula: 
\begin{equation}
\label{vconv}
    v_{\rm{clv}} = \sum_{i=0}^{i=n} c_i \langle\mu\rangle^i
\end{equation}

where n represents the polynomial order and $\langle\mu\rangle$ represents the brightness weighted average value occulted by the planet. The constant offset ($c_0$) in Equation \ref{vconv} is the brightness weighted net convective blueshift integrated over the stellar disk and is removed as we subtract the nightly net out-of-transit convective velocity shift. Full details of this can be found in \citet{cegla2016rossiter}.

\begin{figure}
    \centering
    \includegraphics[width = 0.47\textwidth]{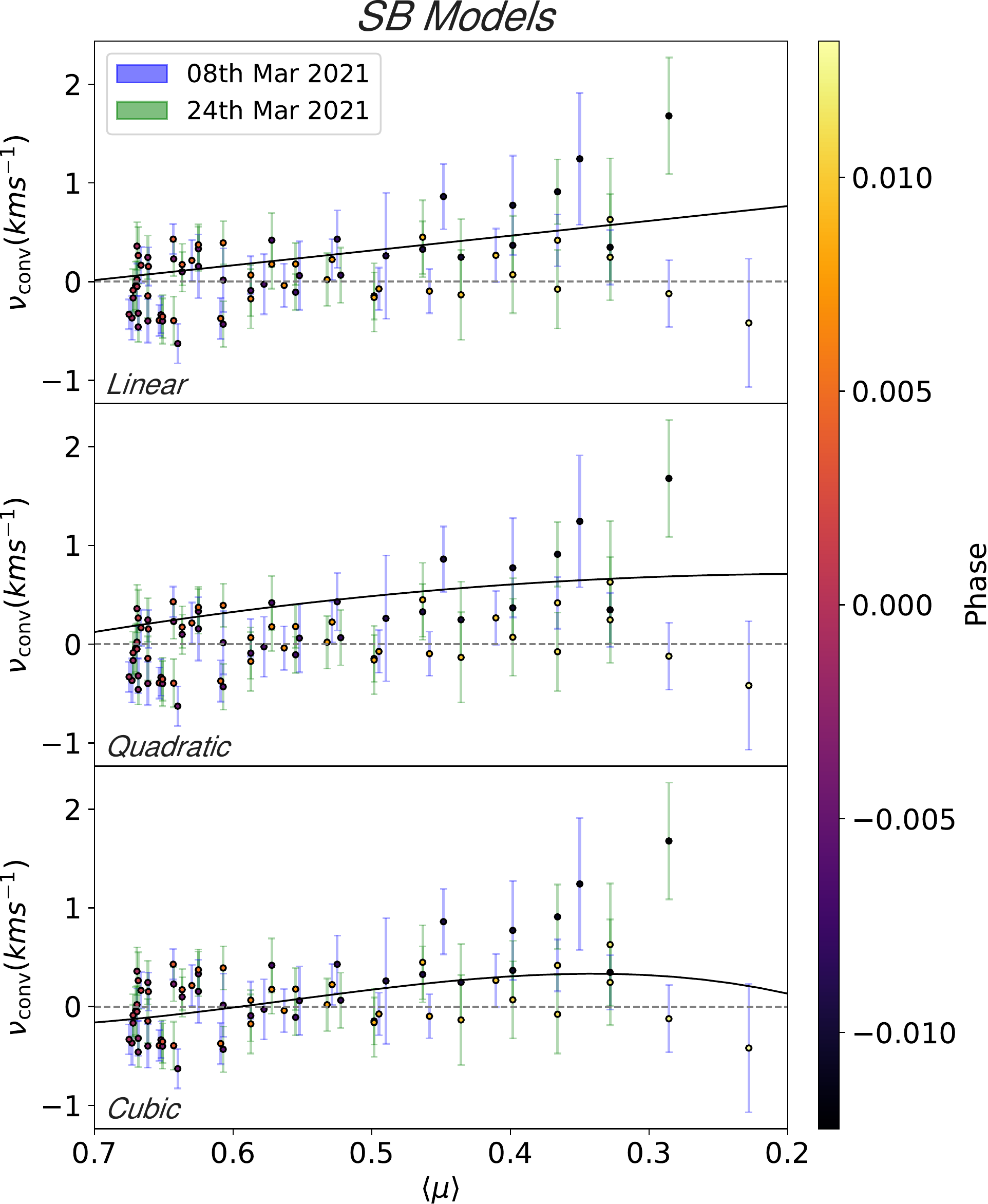}
    \caption{The net convective shifts determined by subtracting the {\it solid body} model fit (which changes slightly when adding in CLV) from the local RVs of the in-transit local CCFs, plotted as a function of stellar disk position behind the planet (brightness weighted $\langle\mu\rangle$). Model fits to the velocities are plotted as linear (top) and quadratic (middle) and cubic (bottom). Horizontal grey dashed lines are plotted at y = 0 to guide the eye.}
    \label{fig:sb_clv}
\end{figure}

The results for the CLV model fits can be found in Table \ref{mcmc_results}, along with the Bayesian Information Criterion (BIC) for each of the models. A difference in BIC of $\sim$6 between models signifies strong evidence of the lower BIC model being the better fit to the data \citep{raftery1995bayesian,lorah2019value}. According to the BICs between the SB models, the best fit to the data is the SB plus a linear CLV which has a lower BIC by $\sim$8, see Figure \ref{fig:sb_clv}. This model gives $v_{\rm{eq}}\sin i_* = 3.05 \pm 0.09$~km s$^{-1}$ (which is still consistent with \citealt{hellier2016wasp}), $\lambda = 167.8 \pm 1.2^{\circ}$ and $c_1 = -1.5 \pm 0.3$~km s$^{-1}$. 

To check the consistency of our results we fit both runs individually for the SB plus linear CLV model. Overall, we find that all fitted parameters for both runs are consistent to within 1$\sigma$. Additionally, we run a SB fit including a white noise term ($\sigma$) to check for additional variability present in the data. Overall, there is a small $\sigma$ value which could potentially be picking up on p-mode oscillations within the local RVs. As a result, we investigated binning the data as a way to effectively average out the p-modes, see \S \ref{sec:binning}. In the SB plus $\sigma$ fit, the \rchi~value is less than one indicating the model is being over-fit to the data causing the $\sigma$ term to be inflated. Therefore, we would not trust this as our best fit model. Additionally, the $\sigma$ term also suggests there may be other contributing noise sources in the data such as magneto convection in the form of granulation/super granulation and/or unaccounted for instrumental effects. 

\subsection{Stellar Differential Rotation}
In the next scenario we fit a model to the local RVs assuming differential rotation (DR) for the star. If DR is present and the planet crosses multiple stellar latitudes, we can determine the true 3D obliquity through disentangling $v_{\rm{eq}}\sin i_*$. Therefore, the model parameters are $\alpha$, $\lambda$, $v_{\rm{eq}}$ and $i_*$ where we assume a differential rotation law derived from the Sun following Equation 8 of \citet{cegla2016rossiter}. Therefore, the stellar rotation velocity at a given position can be expressed as:
\begin{equation}
    v_{\rm{stel}} = x_{\perp} v_{\rm{eq}} \sin i_* (1 - \alpha y\substack{' \\ \perp}^2)
\end{equation}

where $v_{\rm{eq}}$ is the equatorial stellar velocity, $\alpha$ is the differential rotational shear, $i_*$ is the stellar inclination and $x_{\perp}$ and $y\substack{' \\ \perp}$ represent the orthogonal distances from the stellar spin-axis and equator, respectively \footnote{$x_{\perp}$ can be determined by rotating our coordinate system in the plane of the sky by the projected obliquity, $\lambda$. $y\substack{' \\ \perp}$ we then further rotated our coordinate system about the $x_{\perp}$ axis (in the $z_{\perp}y_{\perp}$ plane) by an angle $\beta = \pi/2 − i_*$.} For full details of the coordinate system and equations used we refer the reader to Figure 3 and Section 2.2.1 of \citet{cegla2016rossiter}. We successfully disentangled $v_{\rm{eq}}$ from $i_*$ with results of $7.7 \substack{+1.5 \\ -1.3}$~km s$^{-1}$ and 162.4$\substack{+1.3 \\ -1.2}^{\circ}$, see Table \ref{mcmc_results}. Furthermore, we determined that WASP-131~b crosses a range of latitudes separated by $\sim$60$^{\circ}$ on the stellar surface. Since the projected rotational velocity was determined in the literature from line broadening, it is more appropriate to compare the median product $v_{\rm{eq}}\sin i_*(1 - \alpha y\substack{' \\ \perp}^2)$ to the $v_{\rm{eq}}\sin i_*$ quoted for WASP-131. We find $v_{\rm{eq}}\sin i_*(1 - \alpha y\substack{' \\ \perp}^2) = 2.8 \substack{+0.9 \\ -0.6}$~km s$^{-1}$ which is in agreement with the value of $v_{\rm{eq}}\sin i_* = 3.00 \pm 0.9$~km s$^{-1}$ quoted by \citet{hellier2016wasp}. 

We checked the consistency of our results by fitting both runs individually for the DR alone model. Again, we find the parameters for each of the runs individually are within 1$\sigma$. Similar to the SB model, we ran a DR fit including a white noise term ($\sigma$). For the SB case we found the model was being over-fit to the data, causing an inflation of the $\sigma$ term. For the DR plus $\sigma$ case, we find a similar situation were the \rchi~value is less than one indicating over-fitting of the data. This is precisely what we found for the SB plus $\sigma$ fit, therefore, we would not trust this as our best fit model despite the low BIC value. 

Similar to the SB models, we also accounted for centre-to-limb convective variations and fit the local RVs for both DR and CLV at the same time. This is import because the rotational shear could be on the same order of magnitude as the limb dependant convective variations. For this we tested a linear, quadratic and cubic polynomial as a function of limb angle along with $\alpha$, $\lambda$, $v_{\rm{eq}}$ and $i_*$ for DR rotation where the results can be found in Table \ref{mcmc_results}. For all DR plus CLV model fits, the BIC and \chisq are higher; hence, we concluded the best fit is DR alone. Furthermore, all of the derived polynomial coefficients for the CLV fits are consistent with zero.

In \citet{roguetkern2021drclv} they investigated the optimal parameter space to use the RRM technique to detect DR and CLV on a HD~189733-like system (i.e. a hot Jupiter in a circular orbit around a K-dwarf). To do this they used simulations to explore all possible ranges of $\lambda$, $i_*$ and $b$, producing maps of optimal regions. By placing WASP-131~b with $\lambda \sim 162^{\circ}$, $i_* \sim 40^{\circ}$ (from DR alone model) and $b = 0.7$ (from Table \ref{planetary_properties}) on the heat maps in their Figure 11 we can see that given these conditions we may expect to detect DR and CLV. However, this is with $\alpha$ = 0.2 which is less than a third of what we find for WASP-131 so, a higher $\alpha$ means a better chance at detecting DR.

\begin{figure}
    \centering
    \includegraphics[width = 0.47\textwidth]{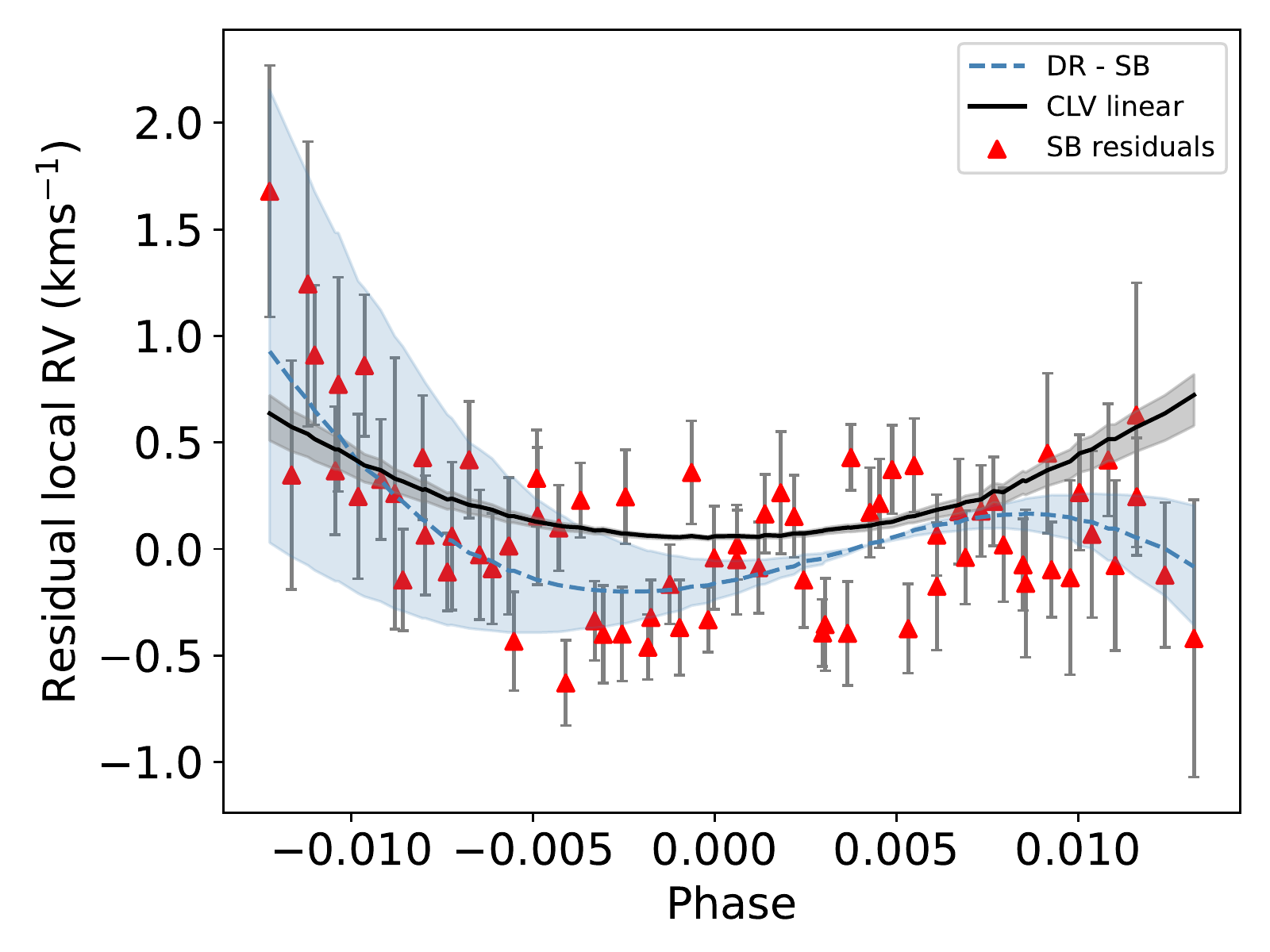}
    \caption{In this plot we compared the linear CLV and differential shear contribution of the local RVs for the un-binned data. The solid black line represents the linear CLV contribution from the SB plus linear CLV model fit, where the grey shaded region represents the errors. The dashed blue line is the SB model subtracted from the DR model, with errors represented as the shaded blue region. Finally, the residuals of the SB model are plotted as red triangles with their corresponding error bars. }
    \label{fig:dr_sb_comp}
\end{figure}

Since SB plus linear CLV was the best fit amongst the SB models and DR alone was the best fit amongst all models, we wanted to make sure we did not confuse CLV and DR. Therefore, we took the SB plus linear CLV model (seen in Figure \ref{local_RVs}) and added Gaussian noise, at the level of the errors, to simulate local RVs. We then fitted the simulated data using an MCMC (as before) testing both a SB plus linear CLV and a DR alone model, finding that the best fit to the data was the SB plus linear CLV model which had a smaller BIC. This then informed us that we are indeed not confusing CLV and DR. Further to this, in Figure \ref{fig:dr_sb_comp} we looked at comparing the linear CLV from the SB fit and the differential shear contribution of the local RVs in the un-binned data. Overall, we find the shape of the SB residuals is similar to both a linear CLV contribution and differential shear contribution. However, there are a few points at ingress and egress (i.e. at the stellar limb) along with several around phase zero which are within the errors and are favourable towards the DR model. Therefore, this explains why the DR model is favoured amongst all models and with more precision in the data it may be possible to pick out DR plus a CLV contribution, especially if more observations are sampled at the limbs of the star. 

\begin{figure}
    \centering
    \includegraphics[width = 0.47\textwidth]{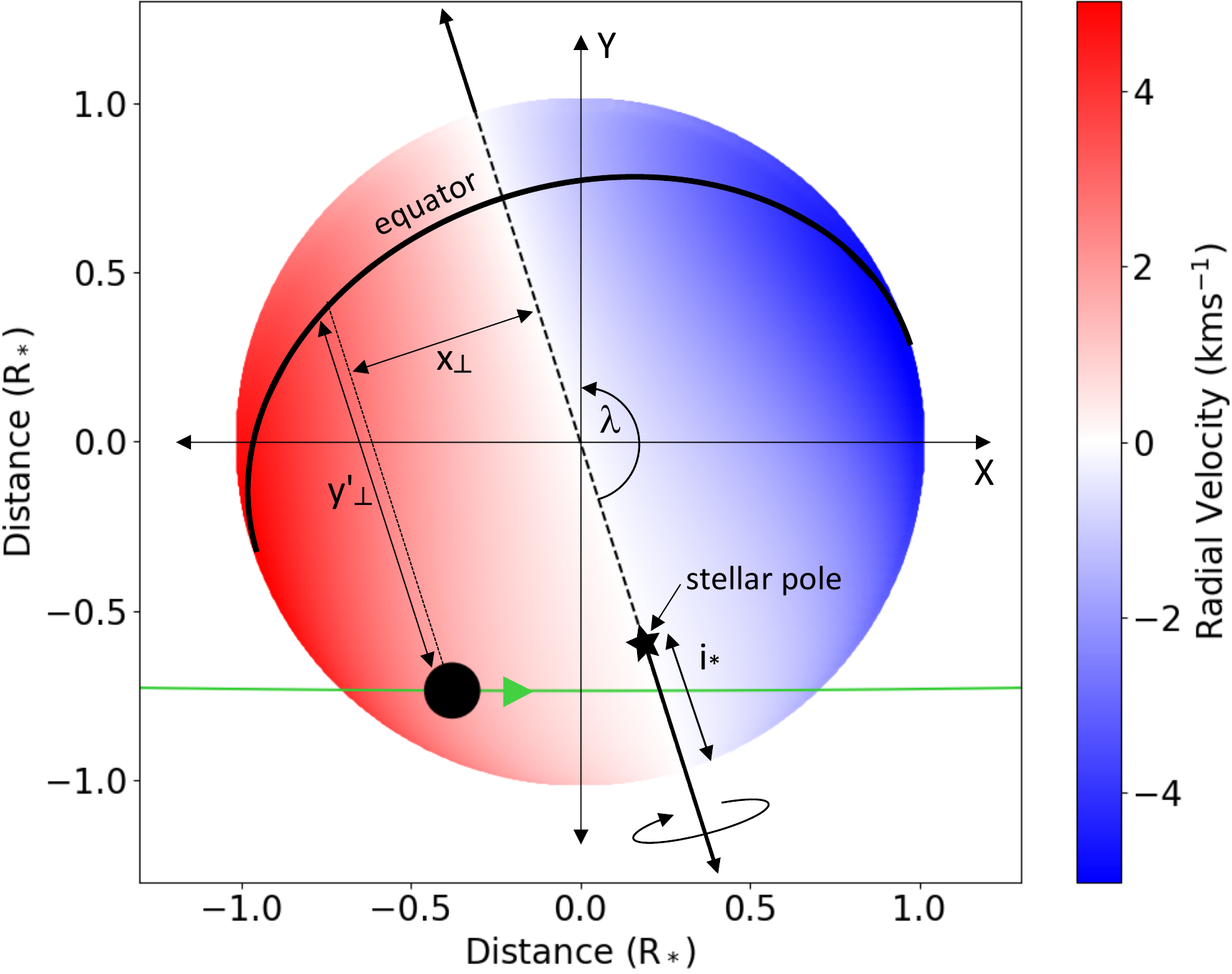}
    \caption{The projection of the WASP-131 system in the plane of sky for the best fit DR model. The entire stellar disk is colour coded according to the radial velocity field. The stellar pole is indicated by a star and the stellar equator a solid black line. The stellar spin axis is represented by a black line, solid when visible and dashed when hidden from view. The black circle represents WASP-131~b to scale where its orbit is a green line.}
    \label{fig:schematic}
\end{figure}

Overall, the best fit model to the data is DR alone which has the lowest BIC amongst both the SB and DR models (excluding those models with the jitter term, where the model overfits the data). Therefore, the final parameters of the system are $v_{\rm{eq}} = 7.7 \substack{+1.5 \\ -1.3}$~km s$^{-1}$, $i_* = 40.9 \substack{+13.3 \\ -8.5}^{\circ}$, $\alpha = 0.61 \pm 0.06$ and $\lambda = 162.4 \substack{+1.3 \\ -1.2}^{\circ}$. Using the relationship, $\psi = \cos(\sin i_* \cos\lambda\sin i_{\rm{p}} + \cos i_*\cos i_{\rm{p}})^{-1} $, we calculated the true 3D obliquity of the WASP-131 system. We find that $\psi = 123.73\substack{+12.82 \\ -8.02}^{\circ}$ meaning the planet is in a polar orbit, see Figure \ref{fig:schematic} for a schematic of the system. 


\subsection{Binned Observations}
\label{sec:binning}
In both the SB and DR models, there is a small $\sigma$ value which could potentially be picking up on p-mode oscillations which are present within the local RVs as a result of stellar variability. In \citet{chaplin2019filtering} they take stellar luminosity, surface gravity and effective temperature to output the ideal exposure time, binning the observations to effectively average out p-mode oscillations. For WASP-131 this worked out as 134.6~mins to suppress the total observed amplitude to a 0.1~ms$^{-1}$ level and 87.3~mins to correspond to the Earth-analogue reflex amplitude of the star. Binning our observations to these times would result in 2 -- 3 data points and so is not a feasible option. Therefore, we referred to ESPRESSO Exposure Time Calculator (ETC) to predict the expected RV precision assuming only photon noise, which is on the 1.5~ms$^{-1}$ level. To reach this, we binned our local CCFs to an exposure time of $\sim$12~mins, averaging out the p-modes to within the uncertainty of the photon noise. This resulted in 19 in-transit data points for both ESPRESSO runs, compared to 39 and 49 for run A and run B, respectively, using un-binned observations.

We refit all SB and DR models to the 12~min binned local RVs, finding the white noise term ($\sigma$) dropped but not completely to zero. This could be a result of remaining correlated red noise, possibly originating from magneto convection such as granulation/super granulation and or residual p-modes. This supports our initial suspicions of p-modes in the un-binned observations. However, given we do not want to lose spatial resolution across the transit chord we do not want to bin the data any further than 12~mins. Overall, all of the fitted parameters for the binned data are consistent with the un-binned fitted parameters to within 1$\sigma$. Therefore, we focus on using the fitted parameters from the un-binned data as our final values and discuss these for the remainder of the paper. 


\subsection{Changes within the CCF profile shape}
Since we fit the local CCFs to extract the stellar velocity behind the planet, we used this fitting to investigate shape changes within the local CCFs. To do this, we analysed the equivalent width (EW), FWHM and contrast as they represent a measure of changes in the area, width, and height of the profile. We fit both a linear and quadratic relationship to each as a function of limb angle and derived the R$^2$ and p-value as a measure of the goodness of fit. The R$^2$ measures the degree to which the data is explained by the model, where a higher value towards 1 indicates a better fit. The p-value then indicates if there is enough evidence that the model explains the data better than a null model (i.e. if the p-value is $<$~0.05, we can reject the null hypothesis). Overall, we found no trend in the EW, FWHM and contrast when fitting both nights of the ESPRESSO observations simultaneously where the distribution is evenly spread about the mean. Similarly for FWHM, fitting each night separately yielded the same result. However, for EW fitting each night separately resulted in a trend being found for run A and nothing for run B. For contrast a trend was found in run B and not run A when fitting each night separately.  

For the EW of run A, R$^2$ = 0.21 and p-value = 0.004 for a linear fit as a function of $\mu$ and R$^2$ = 0.29 and p-value = 0.002 for a quadratic fit, where a quadratic is the better fit to the data. However, only 21\% of the variation in the data is explained by the model. In \citet{doyle2022WASP-166}, we investigated shape changes using EW and FWHM within the local CCFs of WASP-166, a F9 main sequence dwarf with an effective temperature of $T_{\rm{eff}}$ =  6050~K. For WASP-166, we found a quadratic trend was observed in EW and there was a limb-dependent change of $\sim$1~km s$^{-1}$. For WASP-131, in EW we observed a variation of $\sim$2~km s$^{-1}$ where there is more of a spread in the data compared to WASP-166. Overall, an increase in EW could be attributed to be a result of the Fe~I lines being stronger due to the increasing temperature of the lower photosphere with respect to optical depth \citep{beeck2013three}. For the contrast of run B, R$^2$ = 0.21 and p-value = 0.004 for a quadratic fit as a function of $\mu$, where this equates to 7\% of the variation in the data being explained by the model. Overall, this fit is primarily driven by two outliers towards the limb. 

\section{Discussion and Conclusions}
\label{discuss_conclusions}
We have utilised two ESPRESSO transit observations of WASP-131~b to determine, for the first time, the obliquity and conduct a study into stellar surface variability. In addition, we used a number of photometric transit lightcurves from {\tess}, NGTS, EulerCam, WASP and TRAPPIST to update the system properties, including the ephemeris (see Table \ref{planetary_properties}). To determine the obliquity and study stellar surface variability we utilised the Reloaded Rossiter McLaughlin (RRM) technique to determine the local velocities behind WASP-131~b. We then fit the local RVs with various models to account for stellar rotation (solid body and differential) and convective centre-to-limb contributions, see Table \ref{mcmc_results}. Our best fit model to the RVs indicates a detection of stellar surface DR, where we found WASP-131 has a projected obliquity of $\lambda = 162.4\substack{+1.3 \\ -1.2}^{\circ}$, equatorial velocity of $v_{\rm{eq}} = 7.7\substack{+1.5 \\ -1.3}$~km s$^{-1}$, stellar inclination of $i_* = 40.9\substack{+13.3 \\ -8.5}^{\circ}$ and DR shear of $\alpha = 0.61 \pm 0.06$. These values are consistent with WASP-131~b being a misaligned system on a nearly retrograde orbit. 


Furthermore, we were able to determine the true 3D obliquity of the WASP-131 system which is on a polar orbit with $\psi = 123.7\substack{+12.8 \\ -8.0}^{\circ}$. This combined with the high projected obliquity means WASP-131 joins a group of misaligned systems which show a preference for polar orbits. In \citet{albrecht2021preponderance} they determined true 3D obliquity measurements for 57 systems taken from the TEPCAT catalogue \citep{southworth2011homogeneous}, spanning a stellar temperature range of 2500 -- 8500~K (see Figure 3 of the paper for full sample properties), by disentangling $v_{\rm{eq}}\sin i_*$ using stellar rotation periods. They found misaligned systems do not span the full range of obliquities but show a preference for nearly-perpendicular (or polar) orbits with $\psi$ between 80 −- 125$^{\circ}$. There are four theoretical scenarios to explain 3D obliquities near 90$^{\circ}$: (i) tidal dissipation, (ii) Von Zeipel-Kozai-Lidov cycles, (iii) Secular resonance crossing and (iv) Magnetic warping. We will now look at each of these scenarios and discuss whether they could be the cause behind the polar orbit of WASP-131~b.

In \citet{lai2012tidal} they showed tidal dissipation can cause the obliquity to remain at 90$^{\circ}$ rather than damping to 0$^{\circ}$. This happens as a result of damping being dominated  by the dissipation of inertial waves driven in the convective zone by Coriolis forces. WASP-131 has a stellar rotation period which is not greater than double the planetary orbital period for preventing tidal orbital decay. Furthermore, WASP-131~b is considered to have a circular orbit and typically, Von Zeipel-Kozai-Lidov cycles are often used to explain hot Jupiters in highly eccentric orbits. However, tidal dissipation or Von Zeipel-Kozai-Lidov cycles could play a role as the drivers behind the perpendicular/polar orbit of WASP-131~b especially considering the outer companion. 

Considering the outer companion of WASP-131, \citet{petrovich2020disk} proposed secular resonance crossing  where a resonance between the transiting planet and an outer companion occurs as the disk decreases in mass. This resonance can excite the inclination of the inner planet where if the general relativistic procession rate is fast enough, the obliquity is pushed up to 90$^{\circ}$. A very close in companion, which is predicted to be gravitationally bound with a mass 0.62$ \substack{+0.05 \\ -0.04}$~\Msun and separation of $\sim$38~AU, was detected by \citet{bohn2020multiplicity} in the WASP-131 system making this scenario entirely possible. However, \citet{petrovich2020disk} showed this mechanism is more effective for lower-mass, close-orbiting planets, and low-mass, slowly-rotating stars which may rule this mechanism out for WASP-131. 

Finally, magnetic warping can tilt young proto-planetary disks toward a perpendicular orientation, but other mechanisms can counteract this effect so it may not be the leading cause. It is worth remembering that while one of these scenarios may explain our findings it could be a combination of several mechanisms. However, combining our findings with the literature, it is possible tidal dissipation, Von Zeipel-Kozai-Lidov cycles and Secular resonance crossing are among the main drivers. With the presence of an outer companion it is highly likely to be one of these three scenarios responsible for the architecture of the WASP-131 system. Dynamical modelling of the WASP-131 system would help to shed some light on which of these three mechanisms are at play.  

We determined a DR shear of $\alpha = 0.61 \pm 0.06^{\circ}$ for WASP-131. As a result of this, the equator of WASP-131 has a velocity of $v_{\rm{eq}} = 7.7\substack{+1.5 \\ -1.3}$~km s$^{-1}$ where the poles rotate 60\% slower than the equator. In \citet{balona2016differential}, they proposed a relation between DR and the stellar effective temperature along with stellar rotation period. For G and F stars, the shear increases for shorter stellar rotation periods, therefore, the fast stellar rotation period ($P_{\rm{rot}}$ = 11~d: calculated from $v_{\rm{eq}}$ = 7.7~km s$^{-1}$ and $R_*$ = 1.68~\Rsun\ from ARIADNE) of WASP-131 could explain the high derived DR shear. In \citet{reinhold2013rotation} they conducted a study into rotation periods of thousands of Kepler targets spanning a wide range in temperature to search for DR. They found the differential rotational shear weakly depends on temperature for cool stars (3000 -- 6000~K) but above 6000~K, $\alpha$ increases with temperature and the stars in their sample showed no systematic trend and were randomly distributed. WASP-131 has a surface temperature of 5950 $\pm$ 100~K, which combined with the stellar rotation period means this is likely another driver of the high derived DR shear. 

The Fourier transform method to detect DR shear \citep[e.g. see][]{reiners2003rotation} is sensitive to $\alpha > 0.1$ and has been used to detect surface shears as large as 50\% for some A and F stars. In the method by \citet{reinhold2013rotation} they detected DR up to $\alpha < 0.5$, meaning a high solar DR is possible in other stars. Furthermore, they also agreed with \citet{balona2016differential} finding $\alpha$ increases with rotation period for F-G stars also. Overall, DR plays an important role in the generation of magnetic fields within stellar convection zones, and is key for stellar dynamos. Given the high differential rotational shear of WASP-131, it is expected that the star will possess a dynamo mechanism where $\alpha$ will be a key driver in the magnetic field and stellar activity of the star.

In addition to fitting for SB and DR rotation, we also accounted for centre-to-limb convective velocity variations (CLV). The net convective velocity from the centre of the star out to the limb changes due to the convective cells being viewed at different angles from changes in line-of-sight. In both the SB and DR scenarios, we account for CLV by fitting a linear, quadratic and cubic relation as a function of limb angle. Amongst the SB models, the SB plus linear CLV model was preferred where the CLV increases by $\sim$1~km s$^{-1}$ from the centre of the star to the limb linearly (see Figure \ref{fig:sb_clv}), altering the resulting projected obliquity by $\sim$4 degrees when accounted for. However, amongst the DR models, none of the CLV fits were preferred with each of the coefficients effectively zero. 

Finally, we also investigated potential shape changes of the CCFs using EW and FWHM. Overall, there is no trend present in FWHM for either of the observing runs but for EW there is a tentative quadratic trend in run A. This EW trend is similar to that found in \citet{doyle2022WASP-166} of WASP-166 which further solidifies the findings of both \citet{beeck2013three} and \cite{dravins2017spatially, dravins2018spatially} who found similar results in simulated line profiles from state-of-the-art 3D HD simulations.

In \cite{doyle2022WASP-166} we performed a similar analysis on the WASP-166 system where we found SB plus cubic CLV was the best fit to the data. By accounting for CLV we were able to tentatively pull out a DR detection and disentangle $v_{\rm{eq}}$ and $i_*$ putting limits based on if the star is pointing away or towards us. WASP-166 is a bright, $V = 9.36$, F9 main sequence dwarf with an effective temperature of $T_{\rm{eff}}$ = 6050~K, an age of $2.1 \pm 0.9$~Gyr, surface gravity of $log(\rm{g})$ = 4.5 and $\lambda = −15.5\substack{+2.9 \\ -2.8}^{\circ}$. As a reminder, WASP-131 is a G0 main sequence star with V = 10.1, T$_{\rm{eff}}$ = 5950~K, an inflated radius of R$_*$ = 1.70 $\pm$ 0.05 \Msun, age between 4.5 -- 10~Gyr, $log(\rm{g})$ = 3.9 and $\lambda = 162.4\substack{+1.3 \\ -1.2}^{\circ}$. Both stars have p-modes present in the local RVs, for WASP-166 we were able to effectively bin these out as they were on a shorter timescale compared to WASP-131. Both of these stars possess similar stellar properties and so we might expect similar results with regards to the CLV. However, in our best fit model we do not account for CLV. The net convective shifts vary between +0.5~km s$^{-1}$ and -0.5~km s$^{-1}$ where no trend can be identified. This is in stark contrast to the findings of WASP-166 where CLV is characterised by a cubic fit to the net convective velocities which have a velocity of $\sim$ -1 to -2~km s$^{-1}$ at the limb. There are potential degeneracies with the alignment of WASP-131 (which can be seen in the corner plots in Appendix \ref{corner_plots}) which could be causing a null detection of CLV. Furthermore, since SB plus CLV was the best fit amongst the SB models, it may be possible to detect DR plus CLV if we had more precision.  

Overall, we determined the differential rotational shear of WASP-131 and the true 3D obliquitiy of this system for the first time. WASP-131~b joins a group of polar orbiting misaligned planets \citep[see][]{albrecht2021preponderance} which will help shed some light on the processes responsible for their formation and evolution. Dynamical modelling of this system would be interesting to further explore it's formation and evolution especially considering the polar orbit, outer companion and location near the Neptunian desert. Additionally, future observations such as spectropolarimetry or Zeeman Doppler Imaging could be potential ways to investigate the magnetic field and spot activity of WASP-131. This paper forms part of a series where we will use the same ESPRESSO observations to search for various species such as H$_2$O, Na, Li and K in the planetary atmosphere using transmission spectroscopy.

\section*{Acknowledgements}
This work is based on observations made with ESO Telescopes at the La Silla Paranal Observatory under the programme ID 106.21EM. We also include data collected by the {\tess} mission, where funding for the {\tess} mission is provided by the NASA Explorer Program. WASP-South is hosted by the South African Astronomical Observatory where funding comes from consortium universities and from the UK’s Science and Technology Facilities Council. The Euler Swiss telescope is supported by the Swiss National Science Foundation. TRAPPIST is funded by the Belgian Fund for Scientific Research (Fond National de la Recherche Scientifique, FNRS), with the participation of the Swiss National Science Foundation (SNF). We include data collected under the NGTS project at the ESO La Silla Paranal Observatory. The NGTS facility is operated by the consortium institutes with support from the UK Science and Technology Facilities Council (STFC) under grants ST/M001962/1, ST/S002642/1 and ST/W003163/1.

LD and HMC acknowledge funding from a UKRI Future Leader Fellowship, grant number MR/S035214/1. ML acknowledges support of the Swiss National Science Foundation under grant number PCEFP2\_194576. The contributions of ML and AP has been carried out within the framework of the NCCR PlanetS supported by the Swiss National Science Foundation under grants 51NF40\_182901 and 51NF40\_205606. R. A. is a Trottier Postdoctoral Fellow and acknowledges support from the Trottier Family Foundation. This work was supported in part through a grant from the Fonds de Recherche du Qu\'ebec - Nature et Technologies (FRQNT). This work was funded by the Institut Trottier de Recherche sur les Exoplan\'etes (iREx). JSJ greatfully acknowledges support by FONDECYT grant 1201371 and from the ANID BASAL project FB210003.

This project has received funding from the European Research Council (ERC) under the European Union's Horizon 2020 research and innovation programme (project {\sc Spice Dune}, grant agreement No 947634)

\section*{Data Availability}
The {\tess} data are available from the NASA MAST portal and the ESO ESPRESSO data are public from the ESO data archive. CORALIE radial velocities are available through the discovery paper \citet{hellier2016wasp}. The remaining photometry (NGTS, EulerCam etc.) is available as supplementary material online with this paper.


\bibliographystyle{mnras}
\bibliography{WASP-131} 

\begin{thebibliography}{}
\makeatletter
\relax
\def\mn@urlcharsother{\let\do\@makeother \do\$\do\&\do\#\do\^\do\_\do\%\do\~}
\def\mn@doi{\begingroup\mn@urlcharsother \@ifnextchar [ {\mn@doi@}
  {\mn@doi@[]}}
\def\mn@doi@[#1]#2{\def\@tempa{#1}\ifx\@tempa\@empty \href
  {http://dx.doi.org/#2} {doi:#2}\else \href {http://dx.doi.org/#2} {#1}\fi
  \endgroup}
\def\mn@eprint#1#2{\mn@eprint@#1:#2::\@nil}
\def\mn@eprint@arXiv#1{\href {http://arxiv.org/abs/#1} {{\tt arXiv:#1}}}
\def\mn@eprint@dblp#1{\href {http://dblp.uni-trier.de/rec/bibtex/#1.xml}
  {dblp:#1}}
\def\mn@eprint@#1:#2:#3:#4\@nil{\def\@tempa {#1}\def\@tempb {#2}\def\@tempc
  {#3}\ifx \@tempc \@empty \let \@tempc \@tempb \let \@tempb \@tempa \fi \ifx
  \@tempb \@empty \def\@tempb {arXiv}\fi \@ifundefined
  {mn@eprint@\@tempb}{\@tempb:\@tempc}{\expandafter \expandafter \csname
  mn@eprint@\@tempb\endcsname \expandafter{\@tempc}}}

\bibitem[\protect\citeauthoryear{Albrecht, Marcussen, Winn, Dawson  \&
  Knudstrup}{Albrecht et~al.}{2021}]{albrecht2021preponderance}
Albrecht S.~H.,  Marcussen M.~L.,  Winn J.~N.,  Dawson R.~I.,   Knudstrup E.,
  2021, \apjl, 916, L1

\bibitem[\protect\citeauthoryear{Albrecht, Dawson  \& Winn}{Albrecht
  et~al.}{2022}]{albrecht2022stellar}
Albrecht S.~H.,  Dawson R.~I.,   Winn J.~N.,  2022, arXiv preprint
  arXiv:2203.05460

\bibitem[\protect\citeauthoryear{Allard, Homeier  \& Freytag}{Allard
  et~al.}{2012}]{Allard2012}
Allard F.,  Homeier D.,   Freytag B.,  2012, \mn@doi [Philosophical
  Transactions of the Royal Society A: Mathematical, Physical and Engineering
  Sciences] {10.1098/rsta.2011.0269}, 370, 2765

\bibitem[\protect\citeauthoryear{Ara{\'u}jo \& Valio}{Ara{\'u}jo \&
  Valio}{2021}]{araujo2021kepler}
Ara{\'u}jo A.,  Valio A.,  2021, \apjl, 907, L5

\bibitem[\protect\citeauthoryear{{Bailer-Jones}, {Rybizki}, {Fouesneau},
  {Demleitner}  \& {Andrae}}{{Bailer-Jones} et~al.}{2021}]{Bailer-Jones2021}
{Bailer-Jones} C.~A.~L.,  {Rybizki} J.,  {Fouesneau} M.,  {Demleitner} M.,
  {Andrae} R.,  2021, \mn@doi [\aj] {10.3847/1538-3881/abd806}, \href
  {https://ui.adsabs.harvard.edu/abs/2021AJ....161..147B} {161, 147}

\bibitem[\protect\citeauthoryear{Balona \& Abedigamba}{Balona \&
  Abedigamba}{2016}]{balona2016differential}
Balona L.~A.,  Abedigamba O.~P.,  2016, \mnras, 461, 497

\bibitem[\protect\citeauthoryear{Barbary}{Barbary}{2016}]{barbary16sep}
Barbary K.,  2016, \mn@doi [Journal of Open Source Software]
  {10.21105/joss.00058}, 1, 58

\bibitem[\protect\citeauthoryear{Beeck, Cameron, Reiners  \&
  Sch{\"u}ssler}{Beeck et~al.}{2013}]{beeck2013three}
Beeck B.,  Cameron R.~H.,  Reiners A.,   Sch{\"u}ssler M.,  2013, \aap, 558,
  A49

\bibitem[\protect\citeauthoryear{{Bertin} \& {Arnouts}}{{Bertin} \&
  {Arnouts}}{1996}]{bertin1996sextractor}
{Bertin} E.,  {Arnouts} S.,  1996, \mn@doi [\aaps] {10.1051/aas:1996164}, \href
  {https://ui.adsabs.harvard.edu/abs/1996A&AS..117..393B} {117, 393}

\bibitem[\protect\citeauthoryear{Bohn, Southworth, Ginski, Kenworthy, Maxted
  \& Evans}{Bohn et~al.}{2020}]{bohn2020multiplicity}
Bohn A.,  Southworth J.,  Ginski C.,  Kenworthy M.,  Maxted P.,   Evans D.,
  2020, \aap, 635, A73

\bibitem[\protect\citeauthoryear{Bourque et~al.,}{Bourque
  et~al.}{2021}]{matthew_bourque}
Bourque M.,  et~al., 2021, \mn@doi [] {10.5281/zenodo.4556063}

\bibitem[\protect\citeauthoryear{Bourrier, Cegla, Lovis  \&
  Wyttenbach}{Bourrier et~al.}{2017}]{bourrier2017refined}
Bourrier V.,  Cegla H.,  Lovis C.,   Wyttenbach A.,  2017, \aap, 599, A33

\bibitem[\protect\citeauthoryear{{Bryant} et~al.,}{{Bryant}
  et~al.}{2020}]{bryant20multicam}
{Bryant} E.~M.,  et~al., 2020, \mn@doi [\mnras] {10.1093/mnras/staa1075}, \href
  {https://ui.adsabs.harvard.edu/abs/2020MNRAS.494.5872B} {494, 5872}

\bibitem[\protect\citeauthoryear{Cameron et~al.,}{Cameron
  et~al.}{2007}]{cameron2007wasp}
Cameron A.~C.,  et~al., 2007, \mnras, 375, 951

\bibitem[\protect\citeauthoryear{{Castelli} \& {Kurucz}}{{Castelli} \&
  {Kurucz}}{2004}]{Castelli2004}
{Castelli} F.,  {Kurucz} R.~L.,  2004, ArXiv Astrophysics e-prints, \href
  {http://adsabs.harvard.edu/abs/2004astro.ph..5087C} {}

\bibitem[\protect\citeauthoryear{Cegla, Lovis, Bourrier, Beeck, Watson  \&
  Pepe}{Cegla et~al.}{2016a}]{cegla2016rossiter}
Cegla H.,  Lovis C.,  Bourrier V.,  Beeck B.,  Watson C.,   Pepe F.,  2016a,
  \aap, 588, A127

\bibitem[\protect\citeauthoryear{Cegla, Oshagh, Watson, Figueira, Santos  \&
  Shelyag}{Cegla et~al.}{2016b}]{cegla2016modeling}
Cegla H.,  Oshagh M.,  Watson C.,  Figueira P.,  Santos N.~C.,   Shelyag S.,
  2016b, \apj, 819, 67

\bibitem[\protect\citeauthoryear{Chaplin, Cegla, Watson, Davies  \&
  Ball}{Chaplin et~al.}{2019}]{chaplin2019filtering}
Chaplin W.~J.,  Cegla H.~M.,  Watson C.~A.,  Davies G.~R.,   Ball W.~H.,  2019,
  \aj, 157, 163

\bibitem[\protect\citeauthoryear{Claret}{Claret}{2000}]{claret2000new}
Claret A.,  2000, \aap, 363, 1081

\bibitem[\protect\citeauthoryear{{Claret}}{{Claret}}{2004}]{claret2004}
{Claret} A.,  2004, \mn@doi [\aap] {10.1051/0004-6361:20041673}, \href
  {https://ui.adsabs.harvard.edu/abs/2004A&A...428.1001C} {428, 1001}

\bibitem[\protect\citeauthoryear{Collier~Cameron, Donati  \&
  Semel}{Collier~Cameron et~al.}{2002}]{collier2002stellar}
Collier~Cameron A.,  Donati J.-F.,   Semel M.,  2002, \mnras, 330, 699

\bibitem[\protect\citeauthoryear{Collier~Cameron et~al.,}{Collier~Cameron
  et~al.}{2007}]{collier2007efficient}
Collier~Cameron A.,  et~al., 2007, \mnras, 380, 1230

\bibitem[\protect\citeauthoryear{Doyle et~al.,}{Doyle
  et~al.}{2022}]{doyle2022WASP-166}
Doyle L.,  et~al., 2022, \mnras, 516, 298

\bibitem[\protect\citeauthoryear{Dravins, Ludwig, Dahl{\'e}n  \&
  Pazira}{Dravins et~al.}{2017}]{dravins2017spatially}
Dravins D.,  Ludwig H.-G.,  Dahl{\'e}n E.,   Pazira H.,  2017, \aap, 605, A91

\bibitem[\protect\citeauthoryear{Dravins, Gustavsson  \& Ludwig}{Dravins
  et~al.}{2018}]{dravins2018spatially}
Dravins D.,  Gustavsson M.,   Ludwig H.-G.,  2018, \aap, 616, A144

\bibitem[\protect\citeauthoryear{{Gaia Collaboration} et~al.,}{{Gaia
  Collaboration} et~al.}{2016}]{GAIA}
{Gaia Collaboration} et~al., 2016, \mn@doi [\aap]
  {10.1051/0004-6361/201629512}, \href
  {http://adsabs.harvard.edu/abs/2016A%26A...595A...2G} {595, A2}

\bibitem[\protect\citeauthoryear{{Gaia Collaboration}, {Brown}, {Vallenari},
  {Prusti}, {de Bruijne}, {Babusiaux}  \& {Bailer-Jones}}{{Gaia Collaboration}
  et~al.}{2018}]{GAIA_DR2}
{Gaia Collaboration} {Brown} A.~G.~A.,  {Vallenari} A.,  {Prusti} T.,  {de
  Bruijne} J.~H.~J.,  {Babusiaux} C.,   {Bailer-Jones} C.~A.~L.,  2018,
  preprint, \href {http://adsabs.harvard.edu/abs/2018arXiv180409365G} {}
  (\mn@eprint {arXiv} {1804.09365})

\bibitem[\protect\citeauthoryear{Gillon, Jehin, Magain  et~al.}{Gillon
  et~al.}{2011a}]{gillon2011detection}
Gillon M.,  Jehin E.,  Magain P.,   et~al., 2011a.

\bibitem[\protect\citeauthoryear{Gillon et~al.,}{Gillon
  et~al.}{2011b}]{gillon2011wasp}
Gillon M.,  et~al., 2011b, \aap, 533, A88

\bibitem[\protect\citeauthoryear{Hellier et~al.,}{Hellier
  et~al.}{2017}]{hellier2016wasp}
Hellier C.,  et~al., 2017, \mnras, 465, 3693

\bibitem[\protect\citeauthoryear{Husser, von Berg, Dreizler, Homeier, Reiners,
  Barman  \& Hauschildt}{Husser et~al.}{2013}]{Husser2013}
Husser T.-O.,  von Berg S.~W.,  Dreizler S.,  Homeier D.,  Reiners A.,  Barman
  T.,   Hauschildt P.~H.,  2013, \mn@doi [Astronomy {\&} Astrophysics]
  {10.1051/0004-6361/201219058}, 553, A6

\bibitem[\protect\citeauthoryear{{Jenkins} et~al.,}{{Jenkins}
  et~al.}{2016}]{jenkins2016spoc}
{Jenkins} J.~M.,  et~al., 2016, in {Chiozzi} G.,  {Guzman} J.~C.,  eds,
  Society of Photo-Optical Instrumentation Engineers (SPIE) Conference Series
  Vol. 9913, Software and Cyberinfrastructure for Astronomy IV. p. 99133E,
  \mn@doi{10.1117/12.2233418}

\bibitem[\protect\citeauthoryear{Karak, Tomar  \& Vashishth}{Karak
  et~al.}{2020}]{karak2020stellar}
Karak B.~B.,  Tomar A.,   Vashishth V.,  2020, \mnras, 491, 3155

\bibitem[\protect\citeauthoryear{Kitchatinov \& Olemskoy}{Kitchatinov \&
  Olemskoy}{2011}]{kitchatinov2011differential}
Kitchatinov L.,  Olemskoy S.,  2011, \mnras, 411, 1059

\bibitem[\protect\citeauthoryear{{Kurucz}}{{Kurucz}}{1993}]{Kurucz}
{Kurucz} R.,  1993, ATLAS9 Stellar Atmosphere Programs and 2 km/s grid. Kurucz
  CD-ROM No. 13. Cambridge, \href
  {https://ui.adsabs.harvard.edu/abs/1993KurCD..13.....K} {13}

\bibitem[\protect\citeauthoryear{Lai}{Lai}{2012}]{lai2012tidal}
Lai D.,  2012, \mnras, 423, 486

\bibitem[\protect\citeauthoryear{Lendl et~al.,}{Lendl
  et~al.}{2012}]{lendl2012wasp}
Lendl M.,  et~al., 2012, \aap, 544, A72

\bibitem[\protect\citeauthoryear{Lorah \& Womack}{Lorah \&
  Womack}{2019}]{lorah2019value}
Lorah J.,  Womack A.,  2019, Behavior research methods, 51, 440

\bibitem[\protect\citeauthoryear{McLaughlin}{McLaughlin}{1924}]{mclaughlin1924some}
McLaughlin D.,  1924, \apj, 60

\bibitem[\protect\citeauthoryear{Oshagh, Dreizler, Santos, Figueira  \&
  Reiners}{Oshagh et~al.}{2016}]{oshagh2016can}
Oshagh M.,  Dreizler S.,  Santos N.,  Figueira P.,   Reiners A.,  2016, \aap,
  593, A25

\bibitem[\protect\citeauthoryear{Pepe et~al.,}{Pepe
  et~al.}{2014}]{pepe2014espresso}
Pepe F.,  et~al., 2014, Astronomische Nachrichten, 335, 8

\bibitem[\protect\citeauthoryear{Pepe et~al.,}{Pepe
  et~al.}{2021}]{pepe2021espresso}
Pepe F.,  et~al., 2021, \aap, 645, A96

\bibitem[\protect\citeauthoryear{Petrovich, Mu{\~n}oz, Kratter  \&
  Malhotra}{Petrovich et~al.}{2020}]{petrovich2020disk}
Petrovich C.,  Mu{\~n}oz D.~J.,  Kratter K.~M.,   Malhotra R.,  2020, \apjl,
  902, L5

\bibitem[\protect\citeauthoryear{Pollacco et~al.,}{Pollacco
  et~al.}{2006}]{pollacco2006wasp}
Pollacco D.~L.,  et~al., 2006, \pasp, 118, 1407

\bibitem[\protect\citeauthoryear{Queloz, Eggenberger, Mayor, Perrier, Beuzit,
  Naef, Sivan  \& Udry}{Queloz et~al.}{2000}]{queloz2000detection}
Queloz D.,  Eggenberger A.,  Mayor M.,  Perrier C.,  Beuzit J.,  Naef D.,
  Sivan J.,   Udry S.,  2000, \aap, 359, L13

\bibitem[\protect\citeauthoryear{Raftery}{Raftery}{1995}]{raftery1995bayesian}
Raftery A.~E.,  1995, Sociological methodology, pp 111--163

\bibitem[\protect\citeauthoryear{Reiners \& Schmitt}{Reiners \&
  Schmitt}{2002}]{reiners2002feasibility}
Reiners A.,  Schmitt J.~H.,  2002, \aap, 384, 155

\bibitem[\protect\citeauthoryear{Reiners \& Schmitt}{Reiners \&
  Schmitt}{2003}]{reiners2003rotation}
Reiners A.,  Schmitt J.,  2003, \aap, 398, 647

\bibitem[\protect\citeauthoryear{Reinhold, Reiners  \& Basri}{Reinhold
  et~al.}{2013}]{reinhold2013rotation}
Reinhold T.,  Reiners A.,   Basri G.,  2013, \aap, 560, A4

\bibitem[\protect\citeauthoryear{{Ricker} et~al.,}{{Ricker}
  et~al.}{2014}]{ricker2014tess}
{Ricker} G.~R.,  et~al., 2014, in Space Telescopes and Instrumentation 2014:
  Optical, Infrared, and Millimeter Wave. p. 914320 (\mn@eprint {arXiv}
  {1406.0151}), \mn@doi{10.1117/12.2063489}

\bibitem[\protect\citeauthoryear{Roguet-Kern, Cegla  \& Bourrier}{Roguet-Kern
  et~al.}{2022}]{roguetkern2021drclv}
Roguet-Kern N.,  Cegla H.,   Bourrier V.,  2022, Astronomy \& Astrophysics,
  661, A97

\bibitem[\protect\citeauthoryear{Rossiter}{Rossiter}{1924}]{rossiter1924detection}
Rossiter R.,  1924, \apj, 60

\bibitem[\protect\citeauthoryear{Saar \& Donahue}{Saar \&
  Donahue}{1997}]{saar1997activity}
Saar S.~H.,  Donahue R.~A.,  1997, \apj, 485, 319

\bibitem[\protect\citeauthoryear{Schlafly \& Finkbeiner}{Schlafly \&
  Finkbeiner}{2011}]{Schlafly2011}
Schlafly E.~F.,  Finkbeiner D.~P.,  2011, \mn@doi [Astrophysical Journal]
  {10.1088/0004-637X/737/2/103}, 737

\bibitem[\protect\citeauthoryear{Schlegel, Finkbeiner  \& Marc}{Schlegel
  et~al.}{1998}]{Schlegel1998}
Schlegel D.~J.,  Finkbeiner D.~P.,   Marc D.,  1998, \mn@doi [The Astrophysical
  Journal] {10.2478/dema-2014-0019}, pp 525--553

\bibitem[\protect\citeauthoryear{Silva-Valio \& Lanza}{Silva-Valio \&
  Lanza}{2011}]{silva2011time}
Silva-Valio A.,  Lanza A.,  2011, \aap, 529, A36

\bibitem[\protect\citeauthoryear{Smith et~al.,}{Smith
  et~al.}{2020}]{smith2020shallow}
Smith A. M.~S.,  et~al., 2020, Astronomische Nachrichten, 341, 273

\bibitem[\protect\citeauthoryear{Southworth}{Southworth}{2011}]{southworth2011homogeneous}
Southworth J.,  2011, \mnras, 417, 2166

\bibitem[\protect\citeauthoryear{Southworth, Bohn, Kenworthy, Ginski  \&
  Mancini}{Southworth et~al.}{2020}]{southworth2020multiplicity}
Southworth J.,  Bohn A.,  Kenworthy M.,  Ginski C.,   Mancini L.,  2020, \aap,
  635, A74

\bibitem[\protect\citeauthoryear{{Stassun} et~al.,}{{Stassun}
  et~al.}{2019}]{stassun2019ticv8}
{Stassun} K.~G.,  et~al., 2019, \mn@doi [\aj] {10.3847/1538-3881/ab3467}, \href
  {https://ui.adsabs.harvard.edu/abs/2019AJ....158..138S} {158, 138}

\bibitem[\protect\citeauthoryear{{Vines} \& {Jenkins}}{{Vines} \&
  {Jenkins}}{2022}]{ariadne}
{Vines} J.~I.,  {Jenkins} J.~S.,  2022, \mn@doi [\mnras]
  {10.1093/mnras/stac956}, \href
  {https://ui.adsabs.harvard.edu/abs/2022MNRAS.tmp..920V} {}

\bibitem[\protect\citeauthoryear{Virtanen et~al.,}{Virtanen
  et~al.}{2020}]{2020SciPy-NMeth}
Virtanen P.,  et~al., 2020, \mn@doi [Nature Methods]
  {10.1038/s41592-019-0686-2}, \href {https://rdcu.be/b08Wh} {17, 261}

\bibitem[\protect\citeauthoryear{Vogt \& Penrod}{Vogt \&
  Penrod}{1983}]{vogt1983doppler}
Vogt S.~S.,  Penrod G.~D.,  1983, \pasp, 95, 565

\bibitem[\protect\citeauthoryear{{Wheatley} et~al.,}{{Wheatley}
  et~al.}{2018}]{wheatley2018ngts}
{Wheatley} P.~J.,  et~al., 2018, \mn@doi [\mnras] {10.1093/mnras/stx2836},
  \href {https://ui.adsabs.harvard.edu/abs/2018MNRAS.475.4476W} {475, 4476}

\bibitem[\protect\citeauthoryear{Zhao et~al.,}{Zhao
  et~al.}{2022}]{zhao2022expres}
Zhao L.~L.,  et~al., 2022, \aj, 163, 171

\makeatother
\end{thebibliography}

\appendix

\section{Photometric Data}
The photometry used in this paper was uploaded as supplementary material online where an example of the contents for the NGTS simultaneous transits can be seen in Table \ref{tab:phot}.
\begin{table}
	\centering
	\caption{Example table of the NGTS photometry of WASP-131 obtained for this study on the nights 2021 March 08 and 2021 March 24. The full table is available online.}
	\label{tab:phot}
\begin{tabular}{cccc}
    BJD (TDB) &             Flux &  Flux Error & Cam\\
  (-2,450,000) &   &   &  \\ \hline
       9282.61449637 & 1.000652 & 0.006871 & 1 \\
       9282.61461299 & 0.977412 & 0.007051 & 1 \\
       9282.61461935 & 1.003720 & 0.007317 & 1 \\
       9282.61464684 & 0.995525 & 0.006870 & 1 \\
       9282.61470646 & 1.022015 & 0.006940 & 1 \\
       9282.61476346 & 1.021926 & 0.007167 & 1 \\

\hline
	\end{tabular}
\end{table}

\section{MCMC Posterior Probability Distributions}
\label{corner_plots}
\begin{figure*}
    \centering
    \includegraphics[width = 0.70\textwidth]{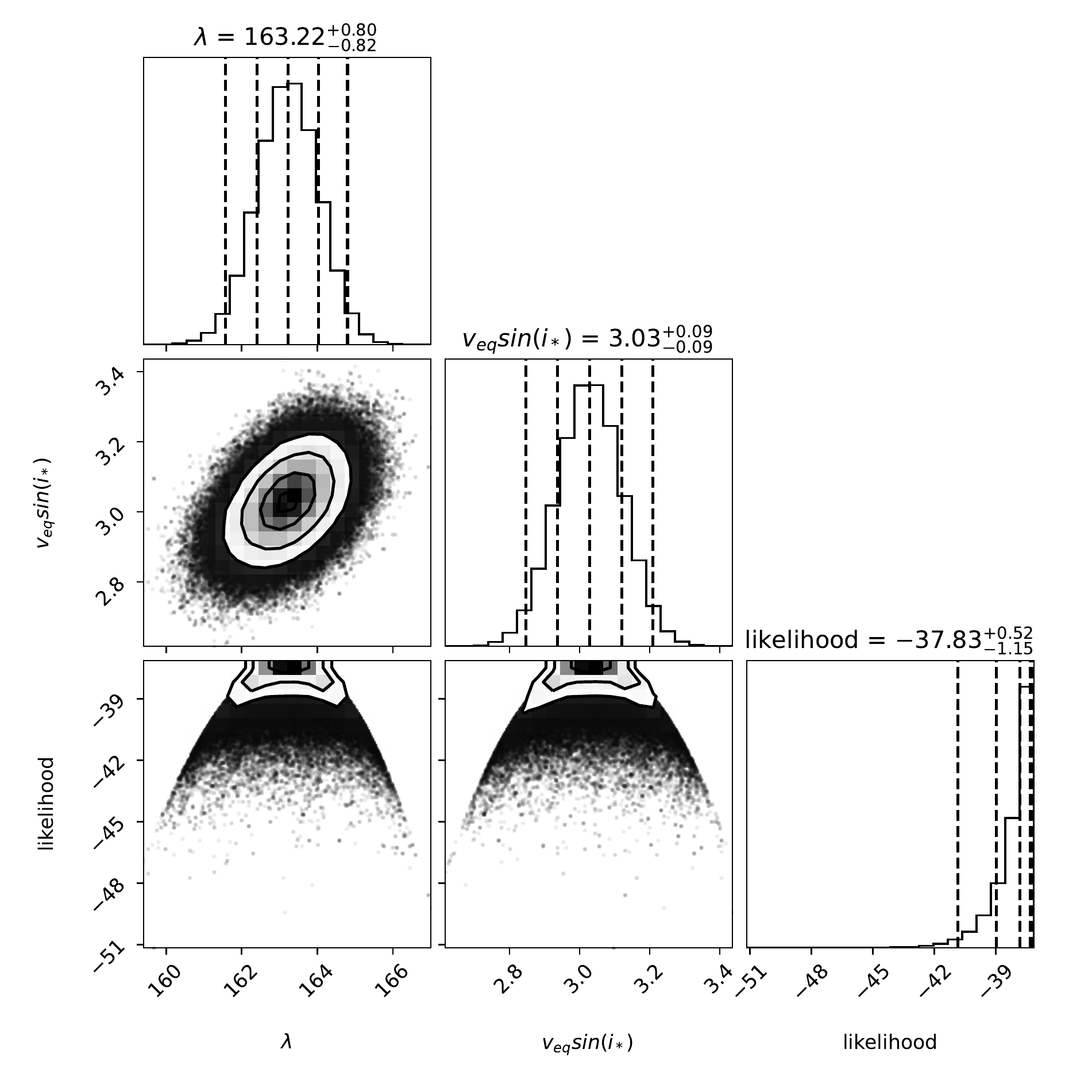}
    \caption{The corner plot of the solid body MCMC showing the one and two dimensional projections of the posterior probability distribution for the parameters. The vertical dashed lines represent the 2.5th, 16th, 50th, 84th and 97.5th percentiles of each parameter.}
    \label{corner:SB}
\end{figure*}

\begin{figure*}
    \centering
    \includegraphics[width = 0.85\textwidth]{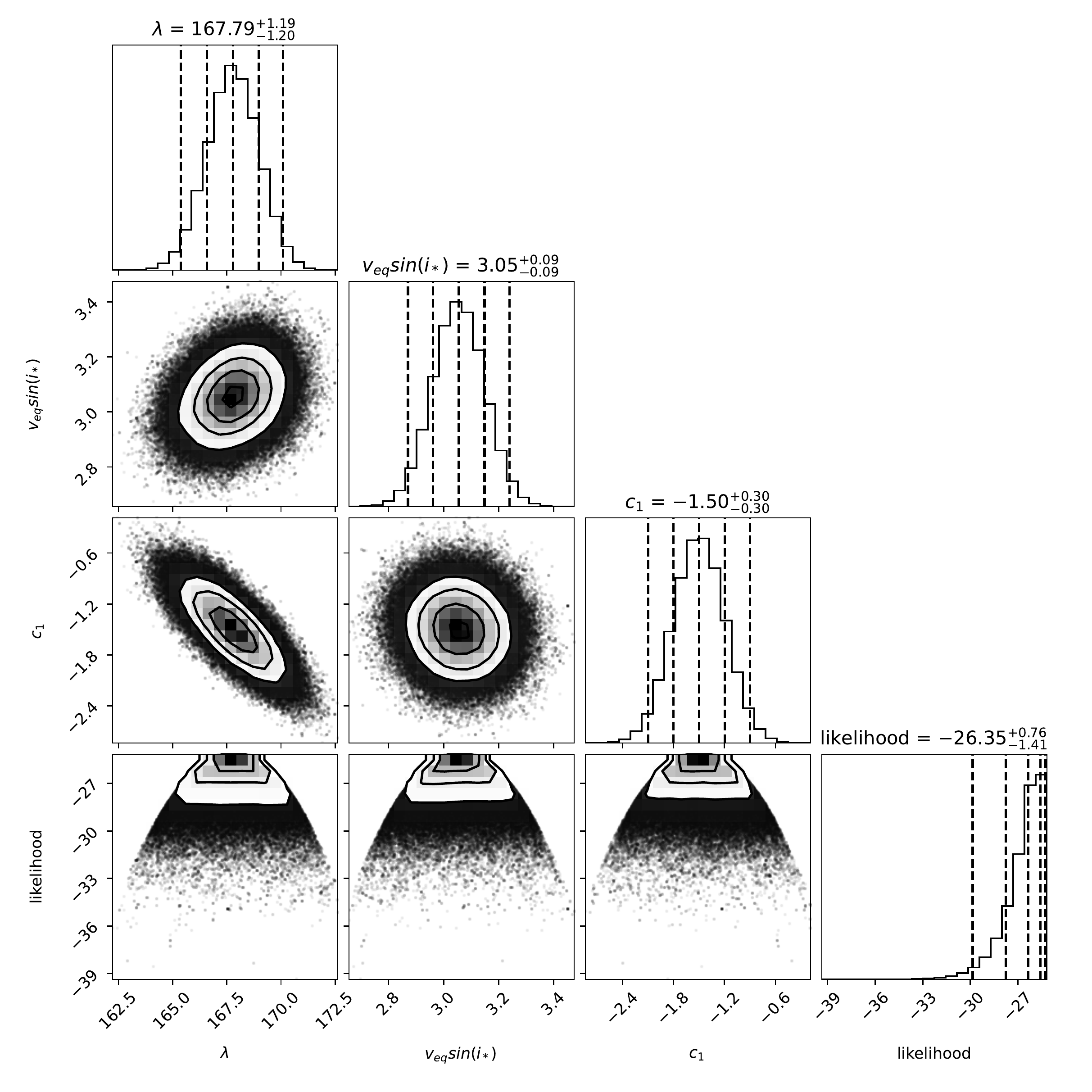}
    \caption{The corner plot of the solid body and linear centre-to-limb convective variations MCMC showing the one and two dimensional projections of the posterior probability distribution for the parameters. The vertical dashed lines represent the 2.5th, 16th, 50th, 84th and 97.5th percentiles of each parameter.}
    \label{corner:SB_clv1}
\end{figure*}

\begin{figure*}
    \centering
    \includegraphics[width = 0.97\textwidth]{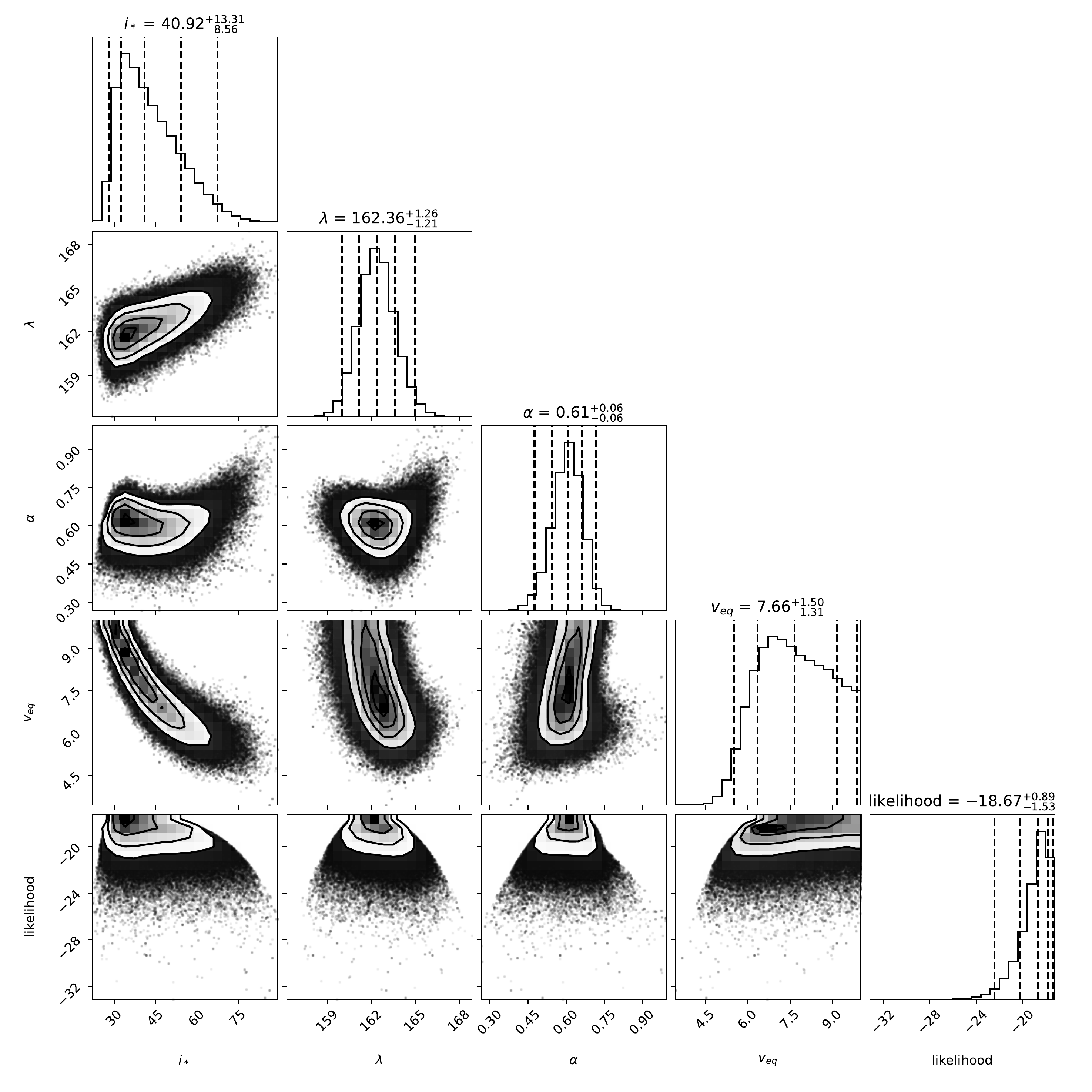}
    \caption{The corner plot of the differential rotation MCMC showing the one and two dimensional projections of the posterior probability distribution for the parameters. The vertical dashed lines represent the 2.5th, 16th, 50th, 84th and 97.5th percentiles of each parameter. There was a boundary placed on $i_*$ at 90$^{\circ}$ due to a low likelihood of a small cluster of outlying points.}
    \label{corner:DR}
\end{figure*}

Here we provide a selection of the corner plots for the MCMC runs showing the one- and two-dimensional projections of the posterior probability distribution for the parameters. These are for SB (Figure \ref{corner:SB}), SB and linear centre-to-limb CLV (Figure \ref{corner:SB_clv1}) and DR (Figure \ref{corner:DR}) models, all fitted to the un-binned observations.

\bsp	
\label{lastpage}
\end{document}